\documentclass[12pt,a4paper]{article}

\usepackage[a4paper, margin=1in]{geometry}
\usepackage[utf8]{inputenc}
\usepackage[T1]{fontenc}
\usepackage{lmodern}
\usepackage[pdftex]{graphicx}
\usepackage{epstopdf}
\usepackage{dsfont}
\usepackage{amssymb,amsmath,amsfonts,amsthm}
\usepackage[round,authoryear,comma]{natbib}
\usepackage{rotating}
\usepackage{epsfig,subfig}
\usepackage{multirow,booktabs}
\usepackage{xcolor}
\usepackage[colorlinks=true,linkcolor=red,urlcolor=blue,citecolor=blue,anchorcolor=blue,pdftex,breaklinks,pdfencoding=auto,psdextra,hypertexnames=false,bookmarksopenlevel=1,bookmarksopen=true]{hyperref}
\graphicspath{{Figures/}{./}}

\usepackage[compact,small]{titlesec}
\usepackage[onehalfspacing]{setspace}
\usepackage{microtype}
\usepackage{authblk}

\makeatletter
\DeclareRobustCommand\citepos
{\begingroup\def\NAT@nmfmt##1{{\NAT@up##1's}}%
	\NAT@swafalse\let\NAT@ctype\z@\NAT@partrue
	\@ifstar{\NAT@fulltrue\NAT@citetp}{\NAT@fullfalse\NAT@citetp}}
\makeatother

\makeatletter
\renewcommand*{\eqref}[1]{\hyperref[{#1}]{\textup{\tagform@{\ref*{#1}}}}}
\makeatother

\newcommand{\eps}{\varepsilon}

\expandafter
\def \expandafter \normalsize \expandafter{\normalsize \setlength \abovedisplayskip{5pt plus 2pt minus 3pt}}
\expandafter
\def \expandafter \normalsize \expandafter{\normalsize \setlength \abovedisplayshortskip{0pt plus 2pt}}
\expandafter
\def \expandafter \normalsize \expandafter{\normalsize \setlength \belowdisplayskip{5pt plus 2pt minus 3pt}}
\expandafter
\def \expandafter \normalsize \expandafter{\normalsize \setlength \belowdisplayshortskip{2pt plus 2pt}}

\begin{document}

\title{The Global Carbon Budget as a cointegrated system\thanks{We thank the editor, Raffaella Giacomini, two anonymous referees, Zack Miller, Eduardo Vera-Vald\'{e}s, Lola Gadea, Chris Smith, Sophocles Mavroeidis, and conference participants at EMCC-VIII in Cambridge 2024 and AWE-V in Aarhus 2024 for helpful comments. We are grateful for financial support from the Independent Research Fund Denmark (IRFD grant number 0219-00001B), Center for Research in Energy:\ Economics and Markets (CORE), the Danish National Research Foundation (DNRF Chair grant number DNRF154), and the Aarhus Center for Econometrics (ACE) funded by the Danish National Research Foundation grant number DNRF186.}}
\author[,1,2]{Mikkel Bennedsen\thanks{\href{mailto:mbennedsen@econ.au.dk}{\texttt{mbennedsen@econ.au.dk}}}}
\author[,2]{Eric Hillebrand\thanks{Corresponding author. \href{mailto:ehillebrand@econ.au.dk}{\texttt{ehillebrand@econ.au.dk}}}}
\author[,1]{Morten \O rregaard Nielsen\thanks{\href{mailto:mon@econ.au.dk}{\texttt{mon@econ.au.dk}}}}

\affil[1]{\small \ Aarhus Center for Econometrics, Aarhus University}
\affil[2]{\small \ Center for Research in Energy, Aarhus University}

\maketitle

\begin{abstract}
The Global Carbon Budget, maintained by the Global Carbon Project, summarizes Earth's global carbon cycle through four annual time series beginning in 1959: atmospheric CO$_2$ concentrations, anthropogenic CO$_2$ emissions, and CO$_2$ uptake by land and by ocean. We analyze these four time series as a multivariate (cointegrated) system. Statistical tests show that the four time series are cointegrated with rank three and identify anthropogenic CO$_2$ emissions as the single stochastic trend driving the nonstationary dynamics of the system. The three cointegrated relations correspond to the physical relations that the sinks are linearly related to atmospheric concentrations and that the change in concentrations equals emissions minus the combined uptake by land and ocean. Furthermore, likelihood ratio tests show that a parametrically restricted error-correction model that embodies these physical relations cannot be rejected on the data. The model can be used for both in-sample and out-of-sample analysis. In an application of the latter, we demonstrate that projections based on this model, using Shared Socioeconomic Pathways scenarios, yield results consistent with established climate science.
\end{abstract}


\newpage

\section{Introduction}

We present a cointegrated vector autoregressive (CVAR) model \citep{Johansen1995,Juselius2006} for the primary time series variables in the Global Carbon Budget (GCB) dataset provided by \cite{GCB2023}. This model enables the statistical estimation of critical parameters of the global carbon cycle and the evaluation of their estimation uncertainty. The model integrates atmospheric carbon dioxide (CO$_2$) concentrations, anthropogenic emissions, and absorption by both the terrestrial biosphere (land sink) and the ocean (ocean sink). Central to the model is the global carbon budget equation: Changes in atmospheric concentrations are equal to the difference between anthropogenic emissions and uptake by the sinks. The model describes the sinks as functions of atmospheric CO$_2$ levels as well as the El Ni\~no-Southern Oscillation (ENSO) and Pacific Decadal Oscillation (PDO) cycles. The budget equation and the dependence of sinks on atmospheric concentrations jointly introduce simultaneity among the GCB variables. Emissions are modeled as a random walk with drift. All variables are trending, and the budget equation as well as the dependence of the sinks on concentrations constitute cointegrating relations.

The approach facilitates a data-driven examination of the global carbon cycle through a compact model incorporating both observational data and outputs from various large-scale Earth system models (ESMs). Historical GCB data are used for parameter estimation via maximum likelihood, and parameter uncertainty is assessed through statistical standard errors. Unlike ESMs or small-scale emulators, the CVAR approach quantifies uncertainty statistically. In contrast to earlier statistical studies of the GCB data set, the CVAR framework allows for formal hypothesis tests of the parametric restrictions motivated by the physical relations.

The Global Carbon Project\footnote{\url{https://www.globalcarbonproject.org}.} maintains an extensive database of time series variables describing the carbon cycle dynamics, aiding in understanding the transfer of anthropogenically emitted CO$_2$ to the atmosphere, oceans, and terrestrial biosphere. These data are updated annually and published in the \emph{Global Carbon Budget} reports. Understanding carbon cycle dynamics is crucial for comprehending the overall climate system and climate change \citep[e.g.][]{AR6chapter5}.

The GCB data have been utilized for various statistical analyses. One research strand investigates whether the CO$_2$ absorption rate of the carbon sinks, measured by the airborne fraction or sink rate, is constant \citep{Raupach2008,Knorr2009,LeQuere2009,Gloor2010,Raupach2014,BHK2019,BHK2024}. Other research uses the GCB data's residual, termed the budget imbalance, to verify the accuracy of reported CO$_2$ emissions by individual nations \citep{Peters2017,Bennedsen2021}. Previous statistical analyses of the GCB data often limit model dimensions to univariate or bivariate settings and rarely consider all GCB variables together. Early studies, such as \cite{Enting1993} and \cite{Parkinson1998}, evaluated parameter uncertainty in global carbon cycle models using statistical methodologies. \cite{BHK2023} specified a multivariate dynamic model of the GCB variables in a state space framework.

In this paper, we leverage the advantages of the CVAR framework that allows for specifying an unrestricted multivariate model that is agnostic about the physical relations of the variables and that nests the restricted, physically motivated model as a special case. Then we conduct a comprehensive (likelihood ratio) hypothesis test for whether the physical restrictions are supported by the data. Since we find that this is the case, we proceed with the restricted model. Earlier applications of the CVAR methodology to different aspects of climate research include \cite{Schmith2012}, \cite{Pretis2020}, and \cite{Castle2024}.

The CVAR model can be used for both in-sample and out-of-sample analysis. In-sample, we demonstrate that the data from the Global Carbon Budget agree with the physical relationships. Out-of-sample, we explore projections of future paths of the global carbon cycle. We find that our projections align well with a common reduced-complexity climate model, in particular for a scenario with strong carbon cycle feedback effects.

The outline of the paper is as follows. In Section~\ref{S:GCBCoint}, we explain the physical relations in the GCB data set and how they motivate a cointegration approach. We specify the restricted, physically motivated model. In Section~\ref{S:test}, we specify the unrestricted model by way of standard cointegration and specification tests. We then conduct a likelihood ratio test of the restricted against the unrestricted model, and we find that the data support the restricted, physically motivated model. We report the estimation results for the restricted model and discuss their physical meaning in the context of the system nature of the model. We also present residual diagnostics. In Section~\ref{S:prog}, we apply the model to explore future projections of the paths of the system using scenarios from the Shared Socioeconomic Pathways (SSP) initiative \citep{Riahi2017}. Section~\ref{S:conc} concludes. An appendix reports additional empirical results and conditional predictive evidence for the model, including validation and forecasting exercises.

\section{Trends and cointegration in the Global Carbon Budget}
\label{S:GCBCoint}

\begin{figure}[t]
\caption{GCB annual time series 1959--2022}
\label{F:GCB_series}
 \includegraphics[width=\textwidth]{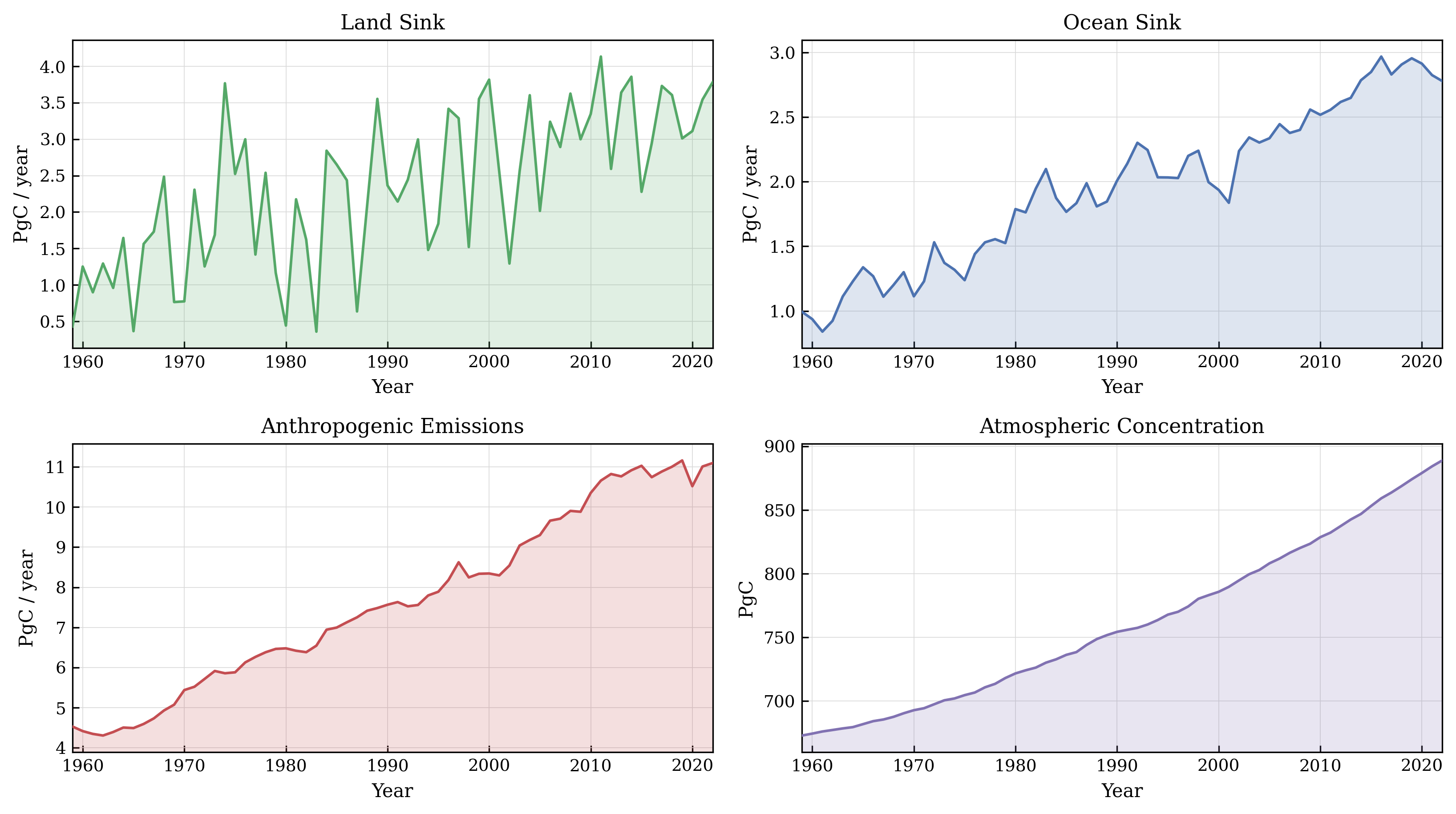}
{\footnotesize Notes: PgC is petagram of carbon (division by 2.12 yields ppm).}
\end{figure}

Figure~\ref{F:GCB_series} displays the time series data set from the Global Carbon Project studied in this paper. The GCB time series are annual from 1959 to 2022, measured in PgC (per year), and obtained from the global file of \cite{GCB2023}.\footnote{Available at \url{https://globalcarbonbudgetdata.org}.} The top left panel in Figure~\ref{F:GCB_series} shows the CO$_2$ uptake by the terrestrial biosphere (land sink), and the top right panel that by the oceans (ocean sink). The bottom left panel shows anthropogenic CO$_2$ emissions, calculated as the sum of fossil fuel and land-use change emissions minus the cement carbonation sink from the global file of \cite{GCB2023}. The bottom right panel shows the ``Keeling curve'' of atmospheric CO$_2$ concentrations.

Figure~\ref{F:GCB_series} shows that all time series are upwards trending. The source of trend in the land and ocean sink dynamics is atmospheric concentrations. \cite{BK1973} and \cite{gifford1993implications} have proposed the most commonly used functions that relate atmospheric CO$_2$ concentrations and land uptake. \cite{BK1973} argue that the linear approximation to their function describes the relation well as long as concentrations do not deviate too far from pre-industrial concentrations. \cite{BHK2023} demonstrate on the GCB data set 1959--2020 that the linear approximation is accurate for both the land and ocean sinks. We therefore specify the sinks equations as
\begin{equation}
\label{E:sinks}
\begin{aligned}
S^L_t &= a_1 + b_1 C_t + X_{1,t}, \\
S^O_t &= a_2 + b_2 C_t + X_{2,t},
\end{aligned}
\end{equation}
where $S^L_t$ and $S^O_t$ are land and ocean sinks, respectively, and $C_t$ is atmospheric concentrations. The deviation processes $X_{i,t} = \phi_i X_{i,t-1} + \eps_{i,t}$, $i=1,2$, where $\eps_{i,t}$ are mutually independent i.i.d.\ zero-mean variables, are specified as AR(1) time series. If $\phi_i = 0$, there is no autocorrelation in the deviations from the trend given by atmospheric concentrations.

\cite{BHK2023} show that anthropogenic emissions $E_t$ are well described by a random walk with drift $d>0$,
\begin{equation}
\label{E:emiss}
E_t = d + E_{t-1} + X_{3,t},
\end{equation}
where $X_{3,t} = \phi_3 X_{3,t-1} + \eps_{3,t}$ and $\eps_{3,t}$ is an i.i.d.\ zero-mean variable. Atmospheric concentrations $C_t$ are given by the \emph{global carbon budget equation},
\begin{equation}
C_t = C_{t-1} + E_t - S^L_t - S^O_t + X_{4,t},\label{E:GCB} 
\end{equation}
where $X_{4,t} = \phi_4 X_{4,t-1} + \eps_{4,t}$ and $\eps_{4,t}$ is an i.i.d.\ zero-mean variable. The budget equation states that changes in atmospheric concentrations, $\Delta C_t$, are given by anthropogenic emissions minus the combined land and ocean uptake plus an error, $X_{4,t}$, which corresponds to the \emph{budget imbalance} variable in the global data file of \cite{GCB2023}. The initial value of $C_t$ in 1959 is 672.87~PgC.

Figure~\ref{F:GCB_series} suggests a possible COVID-19 effect in~2020, where, as a result of the global pandemic, economic activity and consequently emissions dropped substantially. We therefore include a COVID-19 dummy, $D2020_t = I_{\{t=2020\}}$, as an exogenous covariate in the emissions equation.

\begin{figure}[t]
\centering
\caption{Climate covariates 1959--2022}
\label{F:covariates}
\includegraphics[width=\textwidth]{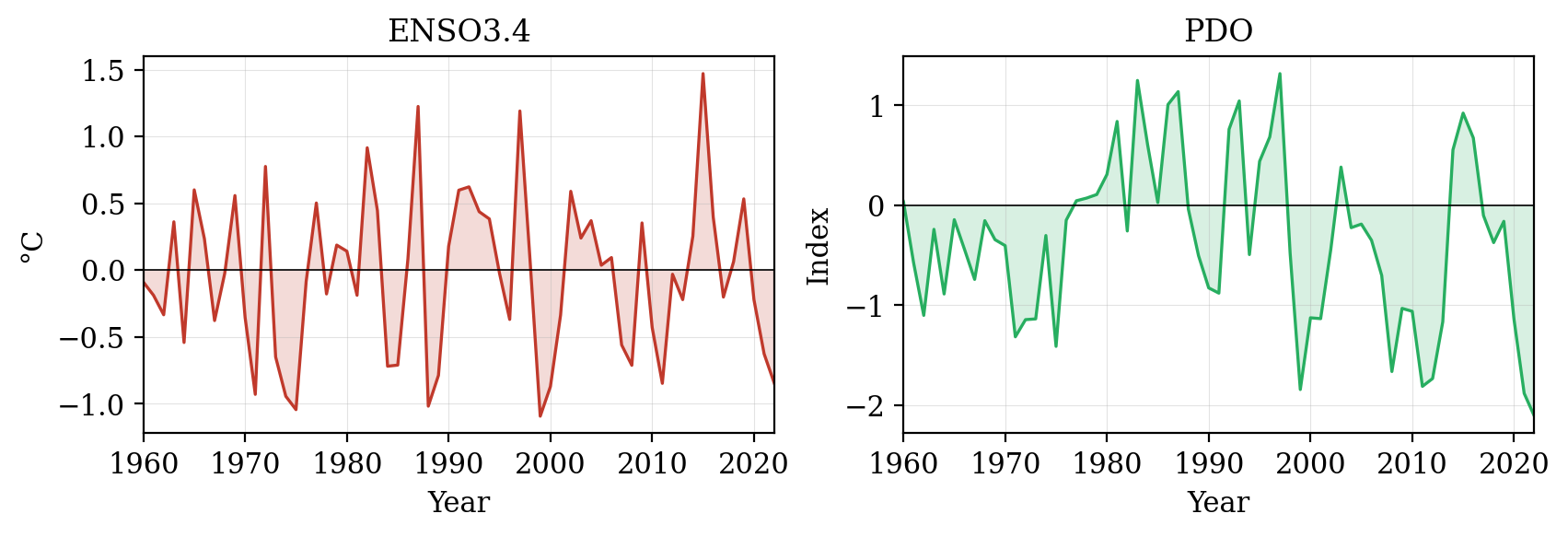}
\end{figure}

Large-scale climate variables have been found to explain part of the variation in the carbon sinks, and via those also some of the variation in concentrations \citep{feely1999influence,haverd2018new,BHK2023}. These climate covariates are stationary, cyclical phenomena, while atmospheric concentrations are nonstationary. Concentrations are therefore by far the dominant explanatory variable in the sinks and central to the cointegrating relations. In this paper, we include two climate covariates, ENSO3.4 and PDO; see Figure~\ref{F:covariates}. ENSO3.4 is a standard measure of ENSO conditions based on sea-surface temperature anomalies in the central equatorial Pacific \citep{Trenberth1997}; we use the monthly Ni\~no~3.4 series provided by the NOAA Physical Sciences Laboratory \citep{Nino34_PSL}, computed from the HadISST dataset \citep{Rayner2003} on a 1981--2010 base period. The PDO is a pattern of decadal sea-surface temperature variability in the North Pacific \citep{Mantua1997}; we use the monthly PDO index distributed by the NOAA National Centers for Environmental Information \citep{PDO_NCEI}, derived from the ERSSTv5 dataset \citep{Huang2017}. Since our analysis is at annual frequency, both indices are aggregated to calendar-year averages of the underlying monthly values.

For the land sink, El Ni\~no conditions are generally associated with reduced CO$_2$ uptake, whereas La Ni\~na conditions are associated with stronger uptake, consistent with the sensitivity of terrestrial carbon uptake to moisture and temperature conditions \citep{haverd2018new}. For the ocean sink, La Ni\~na conditions are associated with reduced uptake because stronger upwelling brings carbon-rich waters to the surface, whereas El Ni\~no conditions weaken this mechanism \citep{feely1999influence}. Hence, we expect the land sink to depend negatively on ENSO3.4 and the ocean sink positively. We include PDO to control for additional low-frequency climate variability beyond ENSO. Specifically, the PDO is a long-lived, ENSO-like mode of North Pacific climate variability, and the carbon-cycle literature attributes part of decadal ocean-sink variability to North Pacific processes. Since the net effects of PDO on global land and ocean uptake are less clear a priori, we do not have similarly strong prior expectations about the signs of its coefficients.

Because emissions are modeled as a random walk \eqref{E:emiss}, and the deviation processes $X_{i,t}$, $i=1,\dots,4$, are all stationary, the model equations in \eqref{E:sinks} and \eqref{E:GCB} suggest that our system of $p=4$ variables is nonstationary I(1) and contains three linear cointegrating combinations that are stationary~I(0). We model the variables collected in $Y_t=[S^L_t, S^O_t, E_t, C_t]'$ in the vector autoregressive (VAR) framework, and the analysis thus fits perfectly in the cointegrated VAR (CVAR) setup; see \citet{Johansen1995,Juselius2006}. Within this systems framework, univariate properties are less relevant (if, e.g., one variable is stationary, then it will appear with a unit vector in the cointegration matrix). For this reason, we do not report univariate unit root or stationarity tests but refer to \cite{BHK2023}.

The CVAR model for a $p$-dimensional vector time series $Y_t$ is given in vector error-correction model (VECM) form as
\begin{equation}
\label{E:VECM_RF}
\Delta Y_t = \mu + \alpha (\beta' Y_{t-1} + \rho t)   + \sum_{j=1}^k \Gamma_j \Delta Y_{t-j} + \Phi Z_t + U_t,
\end{equation}
where the error term $U_t$ is assumed to be $p$-dimensional i.i.d.\ with mean zero and covariance matrix~$\Sigma$. The long-run parameters $\alpha$ and $\beta$ are $p \times r$ matrices with $0 \leq r \leq p$. The rank $r$ is termed the cointegration rank, and the columns of $\beta$ constitute the $r$ cointegration vectors such that $\beta' Y_t$ are the stationary long-run equilibrium relations. The parameters in $\alpha$ are the adjustment parameters that represent the speed of adjustment towards equilibrium for each of the variables. To allow for a deterministic level (mean in the cointegrating relations) and a linear trend in the variables, we have included a so-called unrestricted constant term,~$\mu$, and a restricted trend,~$\alpha \rho t$. Finally, $Z_t$ allows inclusion of exogenous covariates, and the short-run dynamics are governed by the autoregressive parameters $\Gamma_1 , \ldots , \Gamma_k$ with lag-order denoted by~$k$ (that is, there are $k+1$ lags in the VAR representation and equivalently $k$ lagged differences in the VECM representation).

The model specified in \eqref{E:sinks} through \eqref{E:GCB} can be written in structural VAR form as
\begin{align}
\left[\begin{array}{cccc}
1 & 0 & 0 & -b_1 \\
0 & 1 & 0 & -b_2 \\
0 & 0 & 1 & 0 \\
1 & 1 & -1 & 1
\end{array}\right]
\left[\begin{array}{c}
\Delta S^L_t \\ \Delta S^O_t \\ \Delta E_t \\ \Delta C_t
\end{array}\right]
={}& \left[\begin{array}{c}
a_1 (1-\phi_1) \\ a_2 (1-\phi_2) \\ d(1-\phi_3) \\ 0
\end{array}\right]
+ 
\left[\begin{array}{cccc}
c_1 & c_3 & 0 \\
c_2 & c_4 & 0 \\
0 & 0 & \delta \\
0 & 0 & 0 
\end{array}\right]
\left[\begin{array}{c}
ENSO_t \\ PDO_t \\ D2020_t
\end{array}\right]
\notag\\
&+
\left[\begin{array}{cccc}
-(1-\phi_1) & 0 & 0 & b_1(1-\phi_1) \\
0 & -(1-\phi_2) & 0 & b_2(1-\phi_2) \\
0 & 0 & 0 & 0 \\
-(1-\phi_4) & -(1-\phi_4) & 1-\phi_4 & 0
\end{array}\right]
\left[\begin{array}{c}
S^L_{t-1} \\ S^O_{t-1} \\ E_{t-1} \\ C_{t-1}
\end{array}\right] \notag\\
& + \left[\begin{array}{cccc}
0 & 0 & 0 & 0 \\
0 & 0 & 0 & 0 \\
0 & 0 & \phi_3 & 0 \\
0 & 0 & 0 & \phi_4
\end{array}\right]
\left[\begin{array}{c}
\Delta S^L_{t-1} \\ \Delta S^O_{t-1} \\ \Delta E_{t-1} \\ \Delta C_{t-1}
\end{array}\right]
+\left[\begin{array}{c}
\eps_{1,t} \\ \eps_{2,t} \\ \eps_{3,t} \\ \eps_{4,t}
\end{array}\right] ,
\label{E:VECM_w_SOI}
\end{align}
where we have added $Z_t = [ENSO_t, PDO_t, D2020_t]'$ with the relevant coefficient matrix. Left-multiplying \eqref{E:VECM_w_SOI} with the inverse of the leading matrix of contemporaneous relations, we obtain the VECM form \eqref{E:VECM_RF} with
\begin{align}
\alpha &=
\left[\begin{array}{cccc}
-\frac{(1+b_2)(1-\phi_1)}{c} & \frac{b_1(1-\phi_2)}{c} & \frac{b_1(1-\phi_4)}{c} \\
\frac{b_2(1-\phi_1)}{c} & -\frac{(1+b_1)(1-\phi_2)}{c} & \frac{b_2(1-\phi_4)}{c} \\
0 & 0 & 0 \\
\frac{1-\phi_1}{c} & \frac{1-\phi_2}{c} & \frac{1-\phi_4}{c}
\end{array}\right] , \quad
\beta ' =
\left[\begin{array}{cccc}
1 & 0 & 0 & -b_1 \\
0 & 1 & 0 & -b_2 \\
-1 & -1 & 1 & 0
\end{array}\right] , \notag \\
\mu &=
\left[\begin{array}{c}
\frac{a_1}{c} (1-\phi_1)(1+b_2)-\frac{b_1}{c}(a_2(1-\phi_2)-d(1-\phi_3)) \\ \frac{a_2}{c} (1-\phi_2)(1+b_1) -\frac{b_2}{c}(a_1(1-\phi_1)-d(1-\phi_3)) \\ d(1-\phi_3) \\ \frac{1}{c}(d(1-\phi_3)-a_1(1-\phi_1)-a_2(1-\phi_2))
\end{array}\right] , \quad
\rho = \left[\begin{array}{c}
0\\0\\0
\end{array}\right] ,
\notag \\
\Phi &=
\left[\begin{array}{ccc}
\frac{c_1(1+b_2) - b_1 c_2}{c} & \frac{c_3(1+b_2)-b_1 c_4}{c} & \frac{b_1 \delta}{c}\\ 
\frac{c_2(1+b_1) - b_2 c_1}{c} & \frac{c_4(1+b_1)-b_2 c_3}{c} & \frac{b_2 \delta}{c}\\ 
0 & 0 & \delta \\ 
\frac{-c_1-c_2}{c} & \frac{-c_3-c_4}{c} & \frac{\delta}{c}
\end{array}\right] , \quad
\Gamma_1 =
\left[\begin{array}{cccc}
0 & 0 & \frac{b_1\phi_3}{c} & \frac{b_1\phi_4}{c} \\
0 & 0 & \frac{b_2\phi_3}{c} & \frac{b_2\phi_4}{c} \\
0 & 0 & \phi_3 & 0 \\
0 & 0 & \frac{\phi_3}{c} & \frac{\phi_4}{c}
\end{array}\right] , 
\label{E:VECM}
\end{align}
where $c=1+b_1+b_2$, and the error vector is
\[
U_t =
\left[\begin{array}{cccc}
\frac{1+b_2}{c} & \frac{-b_1}{c} & \frac{b_1}{c} & \frac{b_1}{c} \\
\frac{-b_2}{c} & \frac{1+b_1}{c} & \frac{b_2}{c} & \frac{b_2}{c} \\
0 & 0 & 1 & 0 \\
-\frac{1}{c} & -\frac{1}{c} & \frac{1}{c} & \frac{1}{c}
\end{array}\right]
\left[\begin{array}{c}
\eps_{1,t} \\ \eps_{2,t} \\ \eps_{3,t} \\ \eps_{4,t}
\end{array}\right]
=
\left[\begin{array}{c}
\frac{1+b_2}{c} \eps_{1,t} + \frac{b_1}{c}(-\eps_{2,t}+\eps_{3,t}+\eps_{4,t}) \\ \frac{1+b_1}{c} \eps_{2,t} + \frac{b_2}{c}(-\eps_{1,t}+\eps_{3,t}+\eps_{4,t}) \\ \eps_{3,t} \\ \frac{1}{c}(-\eps_{1,t}-\eps_{2,t}+\eps_{3,t}+\eps_{4,t})
\end{array}\right].
\]
The model \eqref{E:VECM_w_SOI} is equivalent to \eqref{E:VECM_RF} with coefficient matrices~\eqref{E:VECM}. The first row of $\beta'$ shows the cointegrating relation between $S^L_t$ and~$C_t$, the second row the cointegrating relation between $S^O_t$ and~$C_t$, and the third row the cointegrating relation of the carbon budget equation between $S^L_t$, $S^O_t$, and~$E_t$. The adjustment coefficients in $\alpha$ show that $S^L_t$, $S^O_t$, and $C_t$ adjust to disequilibrium in all cointegrating relations, whereas $E_t$ does not adjust at all. This shows the special role of emissions as driver of the system, or, in other words, as the only source of deterministic and stochastic trend. It is apparent that the cointegrating rank is three, and the number of common stochastic trends therefore one, and given by the emissions. 

The system \eqref{E:VECM_RF} implied by \eqref{E:VECM} has $k=1$ lagged difference, and the coefficient $\Gamma_1$ of the first lag of differences, $\Delta Y_{t-1} = [\Delta S^L_{t-1}, \Delta S^O_{t-1}, \Delta E_{t-1}, \Delta C_{t-1}]'$, captures the short-run dynamics in stationary first differences of the system variables. It is apparent that autocorrelation in the dynamics of the deviations from the trend in the sinks $S^L_t$ and $S^O_t$, as captured by the autoregressive coefficients $\phi_1$ and $\phi_2$ in~\eqref{E:sinks}, does not play a role in the short-run dynamics of the cointegrated system. In fact, the first two columns of $\Gamma_1$ are zero. The reason can be understood from \eqref{E:VECM_w_SOI}, where accounting for the contemporaneous relation $\Delta S^L_t-b_1\Delta C_t$ on the left-hand side removes all influence of the serial correlation parameter $\phi_1$ from the first differences and fully absorbs it in the dependence on the lagged level~$S^L_{t-1}$. The same holds for~$S^O_t$.

The global carbon budget equation, as a cointegrating relation, implies an influence of the climate covariates in $Z_t$ on~$\Delta C_t$, by way of their influence on the sinks. This leads to a non-zero, but constrained, coefficient on $Z_t$ in the last row of~$\Phi$. The error term $U_t$ is a linear combination of the structural errors~$\eps_t$, by way of the contemporaneous relations. Thus, even though the covariance matrix of $\eps_t$ is diagonal, the covariance matrix of $U_t$ is not diagonal. However, the correlations are purely contemporaneous, and there is no serial correlation.

\section{Estimating and testing the model}
\label{S:test}

In this section, we first determine the lag order. Then we find a benchmark model that is unrestricted except in the cointegrating rank, which we determine using the Johansen trace tests. We then test the restrictions \eqref{E:VECM} implied by the model \eqref{E:VECM_w_SOI} against the unrestricted model, including exogeneity of emissions and the form of the cointegrating relations. Finally, we report estimation results and residual diagnostics for the restricted model~\eqref{E:VECM_w_SOI}.

\subsection{Unrestricted benchmark model}

We first determine the appropriate lag length. The model in \eqref{E:VECM_RF}--\eqref{E:VECM} suggests $k=1$, but of course the unrestricted model \eqref{E:VECM_RF} could have any number of lags $k\geq 0$. Thus, we estimate the unrestricted model \eqref{E:VECM_RF} with full rank and no restrictions on the coefficient matrices for several choices of lag lengths. Because we want to allow for linear trends, we include a constant and a time trend in the unrestricted model. A summary of the results from unrestricted VAR models for $Y_t=[S^L_t, S^O_t, E_t, C_t]'$ are presented in Table~\ref{T:lags} with detailed single-equation residual diagnostics reported in Table~\ref{T:resids} in Appendix~\ref{S:app-empirical}.

\begin{table}[t]
\caption{Lag determination for unrestricted VAR}
\vskip -8pt
\label{T:lags}
\begin{tabular*}{\linewidth}{@{\extracolsep{\fill}}lcccccc}\hline\hline
Lag order & Normality & LB(5) & LB(10) & AIC & LR & $P$-value \\\hline
$k=0$ & 0.631 & 0.584 & 0.621 & 1.482 \\
$k=1$ & 0.327 & 0.157 & 0.431 & 1.774 & 14.232 & 0.581 \\
$k=2$ & 0.080 & 0.004 & 0.148 & 2.135 & \phantom{1}9.947 & 0.869 \\
\hline
\end{tabular*}
\vskip 4pt
{\footnotesize Notes: Normality and LB($j$) are $P$-values for multivariate Jarque-Bera and Ljung-Box Portmanteau tests, the latter to lag~$j$, AIC is the Akaike information criterion, LR is the LR test statistic for exclusion of the $k$-th order lag and $P$-value is its $P$-value. All models are estimated on the common effective sample implied by the maximal lag order, so that the reported criteria and tests are comparable across lag specifications.}
\end{table}

The results in Table~\ref{T:lags} suggest that either $k=0$ or $k=1$ lags would be appropriate. Neither shows signs of misspecification. While the LR test for exclusion of lags suggests that $k=0$ lags could be sufficient, this non-rejection could be caused by the fact that there are many zero entries in the structural $\Gamma_1$ matrix; see~\eqref{E:VECM}. Furthermore, we prefer $k\geq 1$ such that the structural model is nested in the unrestricted model. Thus, we select $k=1$.

\begin{table}[t]
\caption{Trace statistics for cointegrating rank $r$ in the CVAR~\eqref{E:VECM_RF}}
\vskip -8pt
\label{T:trace}
\begin{tabular*}{\linewidth}{@{\extracolsep{\fill}}lccccc}\hline\hline
$r$ & $p-r$ & Eigenvalue & Trace & $P$-value & $P$-value (boot) \\\hline
0 & 4 & 0.4696 & 102.17 & 0.000 & 0.000 \\ 
1 & 3 & 0.3849 & \phantom{1}62.86 & 0.000 & 0.001 \\ 
2 & 2 & 0.2864 & \phantom{1}32.73 & 0.005 & 0.019 \\ 
3 & 1 & 0.1733 & \phantom{1}11.80 & 0.063 & 0.118 \\ 
\hline
\end{tabular*}
\vskip 4pt
{\footnotesize Notes: The model specification has $k=1$ and restricted trend. Bootstrap $P$-values use a residual bootstrap under each rank null.}
\end{table}

A priori, we expect three cointegrating relations to be present in $Y_t$ as given by \eqref{E:sinks} and~\eqref{E:GCB}. Equivalently, emissions are the only source of stochastic trend in the system. Table~\ref{T:trace} confirms that the clear finding from our specification, without imposing any restrictions on the parameters, is a rank of three especially using bootstrap $P$-values for the trace test \citep{CRT2012}. At this point, the statistics cannot tell whether emissions play any specific role.

Consequently, our chosen unrestricted benchmark model is the cointegrated VAR with one lagged difference, cointegration rank three, and including an unrestricted constant term and a restricted trend term, as well as the climate and dummy covariates $Z_t$ as exogenous explanatory variables. The VECM form of the benchmark model is thus given by \eqref{E:VECM_RF} with $p=4$, $k=1$, $r=3$, and $Z_t = [ENSO_t, PDO_t, D2020_t]'$, where $\alpha$ and $\beta$ are of dimension $p\times r = 4\times 3$ and $U_t$ is the $4\times 1$ reduced-form error vector.

Letting $\hat{U}_t$ denote the residuals, the maximized (quasi) log-likelihood function of the model is
\begin{align}
\mathcal{L} &= -\frac{(T-k)p}{2}\log 2\pi - \frac{T-k}{2} \log\det\hat\Sigma -\frac{1}{2}\sum_{t=k+1}^T \hat{U}_t'\hat{\Sigma}^{-1}\hat{U}_t \notag\\
&= -\frac{(T-k)p}{2}\log 2\pi - \frac{T-k}{2} \log\det\hat\Sigma - \frac{(T-k)p}{2},\label{E:llh}
\end{align}
where $T=63$ is the number of observations of $\Delta Y_t$, and 
\[
\hat{\Sigma} = \frac{1}{T-k} \sum_{t=k+1}^T \hat{U}_t\hat{U}_t'
\]
is the estimate of the covariance matrix of the reduced-form errors. Our estimated benchmark model has a numerically maximized log-likelihood of~$-11.373$. Determining the degrees of freedom in the model \eqref{E:VECM_RF} accounts for the $p\times p$ matrix $\alpha \beta'$ of rank $r=3$ with $r(2p-r)=15$ free parameters. In addition, there is the intercept vector $\mu$ of dimension $p$, the trend vector $\rho$ of dimension $r$, and the short-run $p\times p$ coefficient matrix $\Gamma_1$, which together add 23 free parameters. Finally, the $p\times 3$ coefficient matrix on $Z_t$ adds 12 free parameters, for a total of 50 free parameters.

\subsection{Testing the restrictions}

We first test for weak exogeneity and variable exclusion. Testing for weak exogeneity of each variable means testing whether the corresponding row in the $\alpha$ matrix is zero. In that case, the variable does not adjust to disequilibrium in any of the cointegrating relations. This could be either because it is not a relevant variable for the system or because it is a driver of the system. If the variable is not relevant for the cointegrating relations, then the corresponding row in the $\beta$ matrix is zero, i.e., variable exclusion from the cointegration space. In particular, if a variable cannot be excluded, and its $\alpha$ row is zero, then it is causing disequilibrium without adjusting to it, and it is therefore a driver of the system. 

\begin{table}[t]
\caption{LR tests of variable exclusion and weak exogeneity in the CVAR~\eqref{E:VECM_RF}}
\vskip -8pt
\label{T:rf_tests}
\begin{tabular*}{\linewidth}{@{\extracolsep{\fill}}lcccc}\hline\hline
Variable & Exclusion & $P$-value & Weak exogeneity & $P$-value \\\hline
$S^L_t$ & 26.680 & 0.000 & 21.473 & 0.000 \\
$S^O_t$ & 19.595 & 0.000 & 18.213 & 0.000 \\
$E_t$ & \phantom{1}8.152 & 0.043 & \phantom{1}0.379 & 0.945 \\
$C_t$ & \phantom{1}0.545 & 0.909 & 11.309 & 0.010 \\
Trend & \phantom{1}1.692 & 0.639 \\
\hline
\end{tabular*}
\vskip 4pt
{\footnotesize Notes: The model specification has $r=3$, $k=1$, and restricted trend. All LR test statistics are asymptotically $\chi^2$-distributed with three degrees of freedom. The null hypothesis of the exclusion test is a row of zeros in~$\beta$. The null hypothesis of the weak exogeneity test is a row of zeros in~$\alpha$.}
\end{table}

Table~\ref{T:rf_tests} shows the results of LR tests for zero rows in $\beta$ (exclusion) and $\alpha$ (weak exogeneity). The tests for exclusion show that emissions $E_t$ cannot be excluded from the cointegrating relations in the unrestricted model \eqref{E:VECM_RF}, so emissions are important to explain the system dynamics. At the same time, the row in $\alpha$ corresponding to emissions is not distinguishable from zero (weak exogeneity), and therefore emissions do not adjust to disequilibrium. No other variable plays a similar role, so we conclude that emissions are the central driver of the system. Note that, this does not imply that emissions are exogenous with respect to the global economy, or to climate outcomes in a wider sense.

In Table~\ref{T:rf_MA} in Appendix~\ref{S:app-empirical} we present the coefficients in the common trends decomposition of the unrestricted model~\eqref{E:VECM_RF}. The coefficient on emissions is normalized to one, and the remaining coefficients are much smaller in magnitude, which supports the interpretation of emissions as the common trend or driver of the system.

We also note that we cannot reject exclusion of atmospheric concentrations $C_t$ in the reduced-form unrestricted model. However, the inclusion of linear deterministic trends in the sinks in the unrestricted model apparently masks the necessity of $C_t$ in explaining the dynamics and the trending behavior in the sinks. This shows the importance of imposing parameter restrictions for capturing basic physical relations. Specifically, in the structural model \eqref{E:VECM_w_SOI}, the intercept vector and coefficients $\Pi=\alpha\beta'$ and $\Gamma_1$ in \eqref{E:VECM} are such that model \eqref{E:VECM_w_SOI} does not allow for linear deterministic trends in the sinks, so that $C_t$ must be used to explain the trending behavior of the sinks.

The restricted, structural model \eqref{E:VECM_w_SOI} has the parameter vector
\[
\theta = [a_1, a_2, b_1, b_2, c_1, c_2, c_3, c_4, \delta , d, \phi_1, \phi_2, \phi_3, \phi_4 ] ',
\]
which is of dimension~14. The log-likelihood can be evaluated using \eqref{E:llh}, as in the unrestricted model. Model \eqref{E:VECM_w_SOI} is restricted in the system matrices \eqref{E:VECM} in the VECM, that is, in the way the residuals $\hat{U}_t$ are computed. The numerically maximized log-likelihood of model \eqref{E:VECM_w_SOI} is~$-27.967$. Testing the restricted specification \eqref{E:VECM_w_SOI} against the unrestricted specification \eqref{E:VECM_RF} therefore amounts to evaluating the LR test statistic
\[
LR = -2(-27.967-(-11.373)) = 33.188,
\]
which is asymptotically $\chi^2$-distributed with $50-14=36$ degrees of freedom \citep{boswijk2004}. The $P$-value is~0.60, and the restricted model cannot be rejected by the data on the sample period at any standard significance level. We therefore adopt the restricted, structural model \eqref{E:VECM_w_SOI} as our preferred model.

\subsection{Estimation of the restricted model}

\begin{table}[t]
\centering
\begin{minipage}[c]{0.8\linewidth}
\caption{Maximum likelihood estimates of the parameters in model~\eqref{E:VECM_w_SOI}}
\label{T:est}
\vskip -8pt
\begin{tabular*}{\linewidth}{@{\extracolsep{\fill}}lccr}\hline\hline
Parameter & Estimate & Standard error & $t$-statistic \\\hline
$a_1$ & -6.1837 & 1.2607 & -4.9050 \\
$a_2$ & -4.6213 & 0.3316 & -13.9362 \\
$b_1$ & \phantom{-}0.0113 & 0.0017 & 6.7780 \\
$b_2$ & \phantom{-}0.0087 & 0.0004 & 19.9472 \\
$c_1$ & -0.9583 & 0.1759 & -5.4486 \\
$c_2$ & \phantom{-}0.1213 & 0.0234 & 5.1846 \\
$c_3$ & \phantom{-}0.2325 & 0.1280 & 1.8169 \\
$c_4$ & \phantom{-}0.0228 & 0.0193 & 1.1797 \\
$\delta$ & -0.7467 & 0.1607 & -4.6453 \\
$d$ & \phantom{-}0.1182 & 0.0188 & 6.2735 \\
$\phi_1$ & \phantom{-}0.2060 & 0.1097 & 1.8782 \\
$\phi_2$ & \phantom{-}0.5643 & 0.0798 & 7.0743 \\
$\phi_3$ & -0.0961 & 0.1079 & -0.8912 \\
$\phi_4$ & \phantom{-}0.2783 & 0.1212 & 2.2961 \\
\hline
\end{tabular*}
\vskip 4pt
{\footnotesize Notes: Standard errors are computed from the inverse numerical Hessian of the log-likelihood function.}
\end{minipage}
\end{table}

Table~\ref{T:est} reports maximum likelihood estimates of the parameters of model \eqref{E:VECM_w_SOI} together with standard errors. The sinks $S$ (for both $S=S^L$ and $S=S^O$) are measured in PgC and are linear in atmospheric concentrations $C$ in excess of pre-industrial levels, i.e., $S\approx b(C-C_{1750})$. The magnitude of the intercepts is, thus, approximately $a\approx -b C_{1750}$, where the 1750-level atmospheric concentrations are 593~PgC, or about 280~ppm.

The coefficient parameters of the sinks relative to atmospheric concentrations are very similar, and they compare with those found in \cite{BHK2023}. The global carbon budget equation \eqref{E:GCB} states that $C_t - C_{t-1}=E_t - S_t^L -S_t^O + X_t$, where $X_t=X_{4,t}-X_{1,t}-X_{2,t}$ is a stationary error. As a consequence of \eqref{E:sinks}, 
\[
(1+b_1+b_2) C_t - C_{t-1} = E_t - (a_1+a_2) + X_t,
\]
which shows that the dynamics of atmospheric concentrations $C_t$ are determined by the stochastic trend in emissions $E_t$, which propagates in time according to the autoregressive lag polynomial $b(L)=1-(1+b_1+b_2)^{-1}L$. The sum $b_1+b_2$ keeps the dynamics of $C_t$ from a second unit root, in addition to the one in emissions. If the sinks saturate \citep{Canadell2007Saturation,LeQuere2007} and the coefficients $b_1$ and $b_2$ decrease, the dynamics of $C_t$ approach~I(2).

The deviations of the sinks from their trends have different dynamics, as can be seen in Figure~\ref{F:GCB_series}. The trend deviations of the land sink are of higher variance with no visually discernible serial correlation, whereas the deviations of the ocean sink are tighter around the trend with more apparent patterns of serial correlation. This is supported by the estimates of the AR(1) parameters $\phi_1$ (just insignificant at~5\%) and $\phi_2$ (significantly positive). 

The serial correlation in first differences of emissions ($\phi_3$) is indistinguishable from zero, whereas the error in the budget equation has mild serial correlation~($\phi_4$). The drift $d$ in emissions is in line with estimates reported earlier \citep{BHK2023}. The coefficients $c_1$ and $c_2$ on ENSO3.4 in the sinks have the expected sign and are significantly different from zero. The coefficients $c_3$ and $c_4$ on PDO are positive in both sink equations, but they are estimated less precisely, so we do not draw strong conclusions about their effects.

\begin{figure}[t]
\caption{First differences of the system variables and fit of model~\eqref{E:VECM_w_SOI}}
\label{F:VECM}
\includegraphics[width=\textwidth]{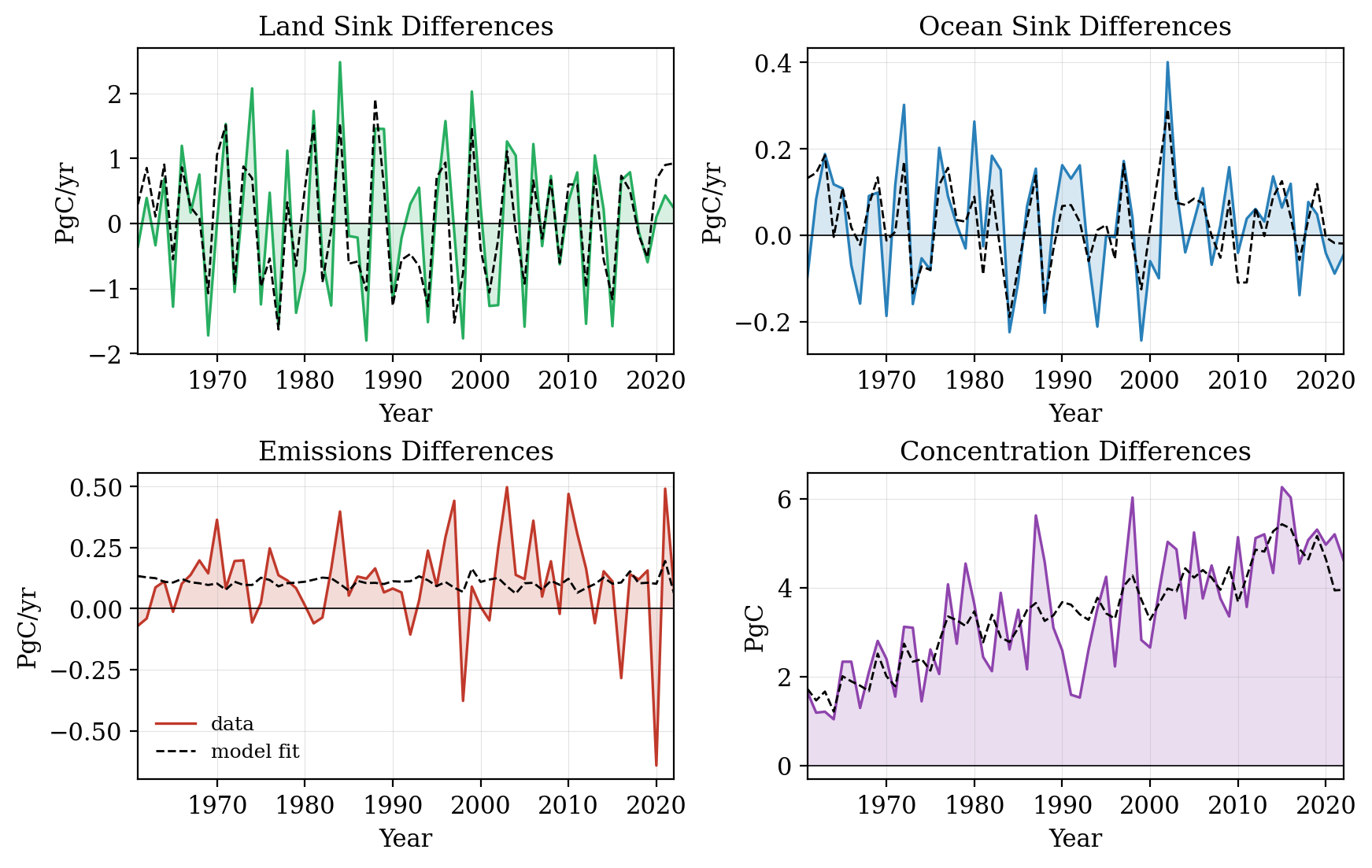}
{\footnotesize Notes: PgC is petagram of carbon (division by 2.12 yields ppm).}
\end{figure}

Figure~\ref{F:VECM} shows the fit of model \eqref{E:VECM_w_SOI} to the data of first differences of the system variables. It can be seen that the model captures variations in the sinks very well. A random walk with drift is a simple model that captures the main role of emissions as driver of the system and source of deterministic and stochastic trend in the cointegrating relations. The lower left panel of Figure~\ref{F:VECM} shows that there is room for improvement in capturing short-run fluctuations of first-differenced emissions, but that is beyond the scope of this paper. In order to explain short-run fluctuations in emissions, macroeconomic data would have to be included in the analysis \citep{BHK2020}.

Figure~\ref{F:VECM} shows that first differences of atmospheric concentrations are upward trending, i.e., that the annual increase in atmospheric concentrations is rising. In the cointegration analysis of the unrestricted model \eqref{E:VECM_RF}, there is an unrestricted linear trend for~$C_t$. 
 In the restricted model \eqref{E:VECM_w_SOI}, in contrast, $\Delta C_t$ is given by the budget equation \eqref{E:GCB} as $\Delta C_t = E_t - S^L_t - S^O_t +X_{4,t}$. Thus, the unrestricted linear trend $dt$ in~$E_t$ and its stochastic trend enter directly into the first differences of concentrations. In addition, the sinks have trending behavior as well, which similarly enters into the first differences of concentrations. The resulting trend in $\Delta C_t$, which is the trend in emissions minus the trends in the sinks, is not equal to zero, and first differences of atmospheric concentrations appear trend-stationary. Evidently, the parametric restrictions are crucial in capturing this important feature of the data.

\begin{table}[t]
\caption{Residual diagnostics for the restricted model~\eqref{E:VECM_w_SOI}
\label{T:resids_restr}}
\vskip -8pt
\begin{tabular*}{\linewidth}{@{\extracolsep{\fill}}lcccccc}\hline\hline
Variable & Std.dev. & Skewness & Kurtosis & Normality & LB(5) & LB(10) \\\hline
$\Delta S^L_t$ & 0.625 & -0.073 & 2.587 & 0.780 & 0.479 & 0.171 \\
$\Delta S^O_t$ & 0.087 & -0.199 & 2.771 & 0.761 & 0.961 & 0.979 \\
$\Delta E_t$ & 0.162 & \phantom{-}0.034 & 3.905 & 0.345 & 0.074 & 0.201 \\
$\Delta C_t$ & 0.662 & -0.147 & 2.278 & 0.456 & 0.940 & 0.103 \\
System &  &  &  & 0.963 & 0.930 & 0.951 \\
\hline
\end{tabular*}
\vskip 4pt
{\footnotesize Notes: System denotes system-wide (multivariate) tests. Std.dev.: standard deviation, Normality: $P$-value of Jarque-Bera test for normality, LB($j$): $P$-value of Ljung-Box test for serial correlation up to lag~$j$.}
\end{table}

Table~\ref{T:resids_restr} shows residual diagnostics for model~\eqref{E:VECM_w_SOI}. There are no signs of misspecification. We note that without the COVID-19 dummy, both the emissions equation and the system tests for Gaussianity would have rejected strongly.

\section{Projections}
\label{S:prog}
In this section, we apply the estimated restricted model to construct out-of-sample projections of the climate variables $C_t$, $S_t^L$, and $S_t^O$, conditional on a path for the exogenous emissions variable $E_t$, over the period 2023--2100. In Section~\ref{S:proj data}, we discuss the exogenous emissions data used for the projections, Section~\ref{S:proj method} lays out the projection methodology, and Section~\ref{S:proj results} presents the projection results. 

Before turning to the scenario projections, we note that Appendix~\ref{app:predictive} provides additional conditional predictive evidence for the restricted model. In a fixed-origin holdout exercise over 2008--2022, the model reproduces the land sink, ocean sink, and atmospheric concentrations reasonably well when conditioned on the realized emissions path, with the climate covariates contributing mainly to the short-run variation in the sink equations. In recursive pseudo-out-of-sample comparisons against less structured benchmark models, the restricted structural model performs particularly well for the land sink and atmospheric concentrations and remains competitive for the ocean sink. Taken together, these historical exercises support the use of the restricted model for the conditional forward-looking analysis conducted in this section.

\subsection{Out-of-sample emissions data}\label{S:proj data}
\begin{figure}[t] 
\caption{Projections of the climate variables under five SSP scenarios}
\label{fig:ECM wFeed RCP}
\includegraphics[width=\textwidth]{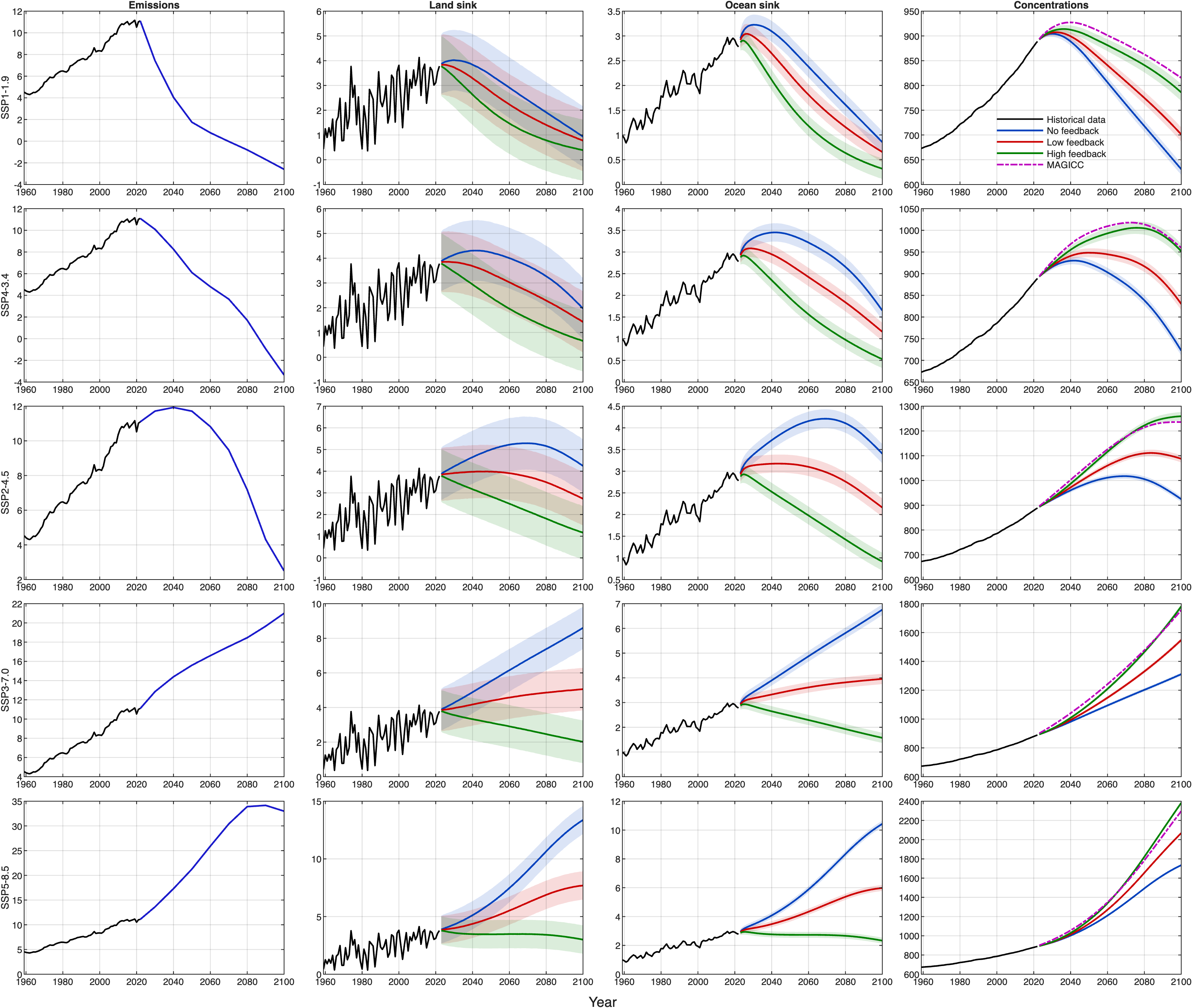} 
{\footnotesize Notes: The five rows correspond to SSP scenarios 119, 434, 245, 370, and 585, respectively. Black: Historical (1959--2022) data. The first column is the CO$_2$ emissions series used as input. Remaining columns present the simulated climate output variables for three different magnitudes of climate feedback, with shaded areas denoting simulated $97.5\%$ and $2.5\%$ pointwise quantiles, and the solid line denoting  $50\%$ pointwise quantiles. Number of simulations: 100,000. Magenta line in right-most column: Projection of atmospheric CO$_2$ concentrations as output from the MAGICC climate model.} 
\end{figure}

We consider emissions trajectories implied by the Shared Socioeconomic Pathways \citep[SSPs;][]{Riahi2017}, as generated by the climate model MAGICC \cite[][]{MAGICC}.\footnote{The MAGICC climate model can be run in a browser at \url{https://live.magicc.org/}.} These trajectories are hypothesized pathways of future anthropogenic CO$_2$ emissions extensively used in the sixth IPCC Assessment Report. SSP scenarios are labeled as ``SSPX-Y'', where ``X'' specifies a certain socioeconomic ``narrative'' used in the construction of the scenario, while ``Y'' describes the level of radiative forcing (given in W/m$^2$) implied by the scenario in the year~2100.\footnote{The conversion from an emissions pathway to a forcing level in 2100 is done using climate models; see \cite{oneill2016} for details.} For instance, \mbox{SSP1-1.9} refers to the SSP scenario in narrative~1, consistent with $1.9$ W/m$^2$ of radiative forcing in the year 2100 above pre-industrial levels.  
Here we do not focus on the various narratives, but only wish to select a number of emissions trajectories that represent a range of possible future emissions pathways. We therefore  select five ``canonical'' SSP scenarios, \mbox{SSP1-1.9}, \mbox{SSP4-3.4}, \mbox{SSP2-4.5}, \mbox{SSP3-7.0}, and \mbox{SSP5-8.5}, as exogenous input into our projections \citep[][]{oneill2016}. These are shown in the left-most panels in Figure~\ref{fig:ECM wFeed RCP}. We refer to \cite{Riahi2017} for further details on the SSP scenarios. 

\subsection{Projection methodology}\label{S:proj method}
Given an exogenous emissions scenario $E_t^*$ for $t = 2023, 2024, \ldots, 2100$, we form the sequence
\[
d_t = E_{t}^\ast-E_{t-1}^\ast,
\]
where 
$E_{2022}^* = E_{2022} = 11.0980$ is set equal to the last in-sample value of the emissions from the GCB data set. The sequence $d_t$ is used as a time-varying drift parameter in the emissions equation of our restricted model and serves to ensure that the model-implied emissions are equal to those of the SSP scenario in the out-of-sample period. In particular, for $t \geq 2023$, we modify \eqref{E:emiss} as follows
\begin{equation}
E_t = d_t + E_{t-1},\label{E:emiss2}
\end{equation}
which, using that $d_t = E_{t}^\ast-E_{t-1}^\ast$, can also be written as
\begin{equation*}
E_t = E_{2022} + \sum_{\tau=2023}^t d_\tau = E_{2022} + \sum_{\tau=2023}^t (E_{\tau}^*-E_{\tau-1}^*) = E_t^\ast,
\end{equation*}
showing that using the time-varying drift $d_t$ and shutting down the disturbance term in the emissions equation~\eqref{E:emiss}, means that the model-implied emissions $E_t$ align with the input SSP trajectory $E_t^\ast$, $t=2023, 2024, \ldots, 2100$.

The sink specifications \eqref{E:sinks} for the restricted model assume that sinks are linear in concentrations. While this assumption is appropriate for the in-sample period \citep[][]{BHK2023}, it may not be an accurate assumption for the out-of-sample period 2023--2100. The reason is that climate change may modify the future sinks/concentration relationships, a phenomenon known as the \emph{climate feedback effect} \citep[e.g.][]{Friedlingstein2015}. Climate feedback effects mean that the sinks likely take up less CO$_2$ over 2023--2100 than the linear model \eqref{E:sinks} would suggest. To capture these feedback effects on the sinks during the out-of-sample period, for $t \geq 2023$ we modify \eqref{E:sinks} as
\begin{equation}
\label{E:sinks2}
\begin{aligned}
S^L_t &= a_{1,t} + b_{1,t} C_t + X_{1,t}, \\
S^O_t &= a_{2,t} + b_{2,t} C_t + X_{2,t},
\end{aligned}
\end{equation}
where, for $j = 1,2$ and $t \geq 2023$,
\begin{equation}
\label{E:slopes}
\begin{aligned}
a_{j,t} = \widehat a_{j} \cdot \exp\left(-\gamma_j  \cdot (t-2022) \right), \\
b_{j,t} = \widehat b_{j} \cdot \exp\left(-\gamma_j  \cdot (t-2022) \right), 
\end{aligned}
\end{equation}
where $\widehat a_{1} = -6.1837$, $\widehat a_2 = -4.6213$, $\widehat b_{1} = 0.0113$, $\widehat b_2 = 0.0087$ are the estimates of $a_{\{1,2\}}, b_{\{1,2\}}$ obtained from the in-sample period 1959--2022, see Table~\ref{T:est}. The parameters $\gamma_{\{1,2\}} \in \mathbb{R}$ determine the degree of feedback in the land and ocean sinks, respectively. When $\gamma_{\{1,2\}} >0$,  the activity of the sinks, as a response to the level of concentrations, weakens as $t$ increases. We note that it is possible that $S_t^{\{L,O\}} <0$ for some $t \geq 2023$, which would correspond to the carbon sink becoming a carbon source. 

There appears to be no consensus in the climate literature on exactly how climate feedback effects modify the sink/concentration relationship in the future. 
Here, we parametrize the feedback effect parameters $\gamma_{\{1,2\}}$ by 
\[
\gamma_{\{1,2\}} = \gamma(p_{\{1,2\}}) = -\frac{\log(1-p_{\{1,2\}})}{28}, 
\]
where $p_{\{1,2\}} \in [0,1]$ denote the relative degree of weakening of the sinks by mid-century, i.e., 28 years after~2022. For instance, $p_1 = 50\%$ would correspond to the case where the land sink activity in 2050 is half of what it would have been in the absence of feedback effects. We consider three different magnitudes of feedback effects specified by $(p_1,p_2) = (0,0)$, $(p_1,p_2) = (25\%,25\%)$, and  $(p_1,p_2) = (50\%,50\%)$. We refer to the first specification as ``no feedback'', the second as ``low feedback'', and the third as ``high feedback''. The no feedback specification corresponds to a situation where the sink parameters $a_{\{1,2\},t}, b_{\{1,2\},t}$ are fixed at the values estimated on the in-sample data, whereas the low (high) feedback specification corresponds to a situation where both the land and ocean sinks have weakened by 25\% (50\%) in~2050.

Using a given SSP scenario for emissions, and substituting \eqref{E:emiss2} and \eqref{E:sinks2} for \eqref{E:emiss} and~\eqref{E:sinks}, respectively, we simulate from the estimated restricted model to generate out-of-sample projections of the climate variables over the period 2023--2100. We generate 100,000 trajectories from the model and report the $2.5\%$, $50\%$, and $97.5\%$ pointwise quantiles. These quantiles reflect sampling uncertainty, and not uncertainty about the emissions scenario, feedback parameters, or parameter estimation.

\subsection{Projection results}\label{S:proj results}

The first column of Figure~\ref{fig:ECM wFeed RCP} shows the exogenous input emissions trajectories for the five SSP scenarios. The second to fourth columns of Figure~\ref{fig:ECM wFeed RCP} present the projected climate variables for these scenarios, derived using the methodology outlined above. The projected CO$_2$ concentrations presented in the right-most column are of particular interest, as they are the main determinants of global temperature changes.

The emissions exhibit substantial variation across scenarios, and the differences directly influence the projected climate outputs. Additionally, the projections vary significantly depending on the degree of climate feedback incorporated. Notably, in the absence of feedback (blue projections), the sinks continue absorbing CO$_2$ according to \eqref{E:sinks}, resulting in substantially lower CO$_2$ concentrations compared to cases with climate feedbacks included.

The magenta dashed lines in the last column of Figure~\ref{fig:ECM wFeed RCP} represent the concentration pathway projected by the MAGICC climate model. These projections align closely with those from the high feedback specification of our restricted model. This alignment is significant for two reasons. First, despite the more complex equations used in MAGICC, our model produces comparable results for all five SSP scenarios studied here, underscoring its capability to replicate projections from detailed climate models. Second, this similarity arises only under the high feedback specification, where both sinks weaken by $p_1 = p_2 = 50\%$ in~2050. This suggests that consistency with the historical data record, on which our model is estimated, requires large climate feedback effects to materialize between 2023 and~2100.

For context, the third IPCC Assessment Report estimated mid-century feedback effects of 21\%--43\% for the land sink and 6\%--25\% for the ocean sink \citep[][]{IPCC2001_3rd}. However, more recent studies indicate that these estimates may be conservative, underestimating the potential magnitude of future feedback effects \citep[][]{Friedlingstein2015}.

\section{Conclusions}
\label{S:conc}

In this paper, we have specified a cointegrated vector autoregressive (CVAR) model for the global variables in the Global Carbon Budget data set. In contrast to earlier comprehensive statistical models of the GCB data, the CVAR approach allows for formal hypothesis tests of the physically motivated functional forms of the land and ocean sinks, the emission dynamics, and the budget equation. The global carbon budget equation and the dependence of the sinks on atmospheric CO$_2$ concentrations imply simultaneous relations between the variables, and they constitute cointegrating relations at the same time. We specified both a restricted and physically motivated model as well as an unrestricted and physically agnostic model that nests the restricted model as a special case. A likelihood ratio test showed that the physically motivated, restricted model is supported by the data. We discussed the estimation results in light of the system nature of the variables. We then used the restricted model to explore future projections of the path of the global carbon cycle using SSP scenarios. 
Recent discussion in the literature \citep{Canadell2007Saturation,LeQuere2007} has suggested possible (partial) saturation in the land and ocean sinks in the near future. In the context of our modeling framework, this would constitute a nonlinear, time-dependent cointegrating relationship. Analyzing the GCB data in a model that allows for time-dependence and nonlinearity in the relation between the sinks and atmospheric concentrations is an interesting topic for future research.

\newpage
\clearpage

\appendix

\section{Supplementary empirical results}
\label{S:app-empirical}

This appendix contains additional empirical results. Table~\ref{T:resids} presents residual diagnostics equation-by-equation for the unrestricted VAR models. Table~\ref{T:rf_MA} presents the coefficients $\tilde\beta_{\perp} = \hat\beta_{\perp} (\hat\alpha_{\perp}' \hat\Gamma \hat\beta_{\perp})^{-1}$ and $\tilde\alpha_{\perp} = \hat\Gamma \hat\alpha_{\perp}$ of the permanent part of the permanent-transitory decomposition of~$Y_t$. These are based on premultiplying $Y_t$ by
\begin{equation*}
\beta_{\perp}(\alpha_{\perp}'\Gamma\beta_{\perp})^{-1}\alpha_{\perp}'\Gamma + \Gamma^{-1}\alpha(\beta'\Gamma^{-1}\alpha)^{-1}\beta'=I_p,
\end{equation*}
where the first term yields the permanent component and the second term yields the transitory component; see \citet{GG1995} and \citet[Corollary~4.4]{Johansen1995}. We note that the variables are measured on different scales, which explains the large differences in the values of the entries in $\tilde\beta_{\perp}$.

\begin{table}[h]
\caption{Residual diagnostics for unrestricted VARs}
\vskip -8pt
\label{T:resids}
\begin{tabular*}{\linewidth}{@{\extracolsep{\fill}}lcccccc}\hline\hline
Variable & Std.dev. & Skewness & Kurtosis & Normality & LB(5) & LB(10) \\\hline
\multicolumn{7}{c}{$k=0$ lags} \\
$S^L_t$ & 0.606 & -0.166 & 2.794 & 0.824 & 0.368 & 0.089 \\
$S^O_t$ & 0.075 & -0.195 & 3.415 & 0.662 & 0.762 & 0.880 \\
$E_t$ & 0.147 & -0.229 & 3.602 & 0.483 & 0.140 & 0.186 \\
$C_t$ & 0.864 & -0.126 & 3.023 & 0.922 & 0.946 & 0.542 \\\hline
\multicolumn{7}{c}{$k=1$ lag} \\
$S^L_t$ & 0.579 & -0.027 & 2.920 & 0.988 & 0.470 & 0.410 \\
$S^O_t$ & 0.073 & -0.144 & 3.273 & 0.819 & 0.884 & 0.894 \\
$E_t$ & 0.144 & -0.441 & 4.084 & 0.084 & 0.049 & 0.171 \\
$C_t$ & 0.809 & -0.207 & 2.946 & 0.800 & 0.402 & 0.174 \\\hline
\multicolumn{7}{c}{$k=2$ lags} \\
$S^L_t$ & 0.562 & -0.237 & 2.811 & 0.718 & 0.427 & 0.679 \\
$S^O_t$ & 0.071 & -0.464 & 3.386 & 0.278 & 0.646 & 0.828 \\
$E_t$ & 0.143 & -0.446 & 4.360 & 0.035 & 0.018 & 0.106 \\
$C_t$ & 0.799 & -0.348 & 3.130 & 0.529 & 0.133 & 0.076 \\\hline
\end{tabular*}
\vskip 4pt
{\footnotesize Notes: Std.dev.: standard deviation, Normality: $P$-value of Jarque-Bera test for normality, LB($j$): $P$-value of Ljung-Box test for serial correlation up to lag~$j$. The models are \eqref{E:VECM_RF} with unrestricted coefficients and full rank, including a constant and time trend. All models are estimated on the common effective sample implied by the maximal lag order, so that the reported criteria and tests are comparable across lag specifications.}
\end{table}

\begin{table}[hb]
\centering
\begin{minipage}{0.65\linewidth}
\caption{Common trends decomposition of the CVAR~\eqref{E:VECM_RF}}
\vskip -8pt
\label{T:rf_MA}
\begin{tabular*}{\linewidth}{@{\extracolsep{\fill}}lcc}
\hline\hline
Variable & $\tilde\alpha_{\perp}$ & $\tilde\beta_{\perp}$ \\
\hline
$S^L_t$ & \phantom{-}0.000 & \phantom{-1}0.557 \\
$S^O_t$ & -0.156 & \phantom{-1}0.093 \\
$E_t$   & \phantom{-}1.000 & \phantom{-1}0.699 \\
$C_t$   & -0.018 & -17.940 \\
\hline
\end{tabular*}
\vskip 4pt
{\footnotesize Notes: The model specification has $r=3$, $k=1$, an unrestricted constant, and a restricted trend.}
\end{minipage}
\end{table}

\newpage

\section{Supplementary conditional predictive evidence}
\label{app:predictive}

This appendix reports additional conditional predictive evidence for the restricted structural model used in the paper. The exercises are designed to assess the model in the same conditional sense in which it is used for the scenario projections in Section~\ref{S:prog}. First, we ask whether the model can reproduce the final part of the historical sample when conditioned on realized emissions and, when relevant, realized climate covariates. Second, we compare its recursive pseudo-out-of-sample conditional forecasting performance with less structured benchmark models. Throughout, the objects of interest are conditional forecasts of the land sink~$S_t^L$, the ocean sink~$S_t^O$, and atmospheric concentrations~$C_t$, given the realized path of anthropogenic emissions and, when included, the realized path of the stationary climate covariates $Z_t = [ENSO_t,PDO_t]'$.

\subsection{Validation}
\label{app:validation}
\label{app:validation_goal}

The validation exercise asks whether the restricted model can reproduce the final observed years of the sample under the same type of conditioning used in the forward-looking projections. In the main text, projections are constructed conditional on an exogenous emissions path. Here we ask the corresponding retrospective question: if the realized path of emissions, and possibly the realized path of the climate covariates, had been known, how well would the model have reproduced the final observations for $S_t^L$, $S_t^O$, and~$C_t$?

We conduct a fixed-origin holdout exercise with a 15-year holdout block. This choice is long enough to assess medium-run conditional dynamics while leaving a reasonably large estimation sample. Let $T$ denote the final observed year and set $H=15$. The model is estimated on the truncated sample ending in year $T-H$ and then projected over the holdout period $T-H+1,\ldots,T$. The projections are conditional on the realized emissions path over the holdout period and, in the climate-augmented specification, also on the realized paths of~$Z_t$.  

\subsubsection{Models}
\label{app:validation_models}

We consider two versions of the restricted structural model.

\paragraph{Restricted model with climate covariates.}
The first specification is the restricted structural model including the stationary climate variables,~$Z_t$. The land and ocean sink equations include contemporaneous climate terms, while atmospheric concentrations evolve subject to the carbon-budget restriction. Forecasts are conditional on realized emissions and realized~$Z_t$. This specification should therefore be viewed as an additional diagnostic of the structural model's conditional mapping from emissions and climate inputs into sinks and atmospheric concentrations.

\paragraph{Restricted model without climate covariates.}
As a comparison, we also report projections from the same restricted structural system with the coefficients on $Z_t$ set to zero. This specification preserves the carbon-cycle restrictions but suppresses the direct climate-covariate channel in the sink equations. It is also the closer historical analogue to the scenario projections in Section~4, which are conditional on an exogenous emissions path but not on realized future~$Z_t$.

For both model specifications, we report pointwise 95\% predictive intervals obtained by simulation, conditional on the realized exogenous paths. These intervals reflect innovation uncertainty conditional on the realized emissions and, for the climate-augmented specification, climate inputs.

\subsubsection{Results}
\label{app:validation_results}

\begin{figure}[t]
\caption{Validation exercise over the 2008--2022 holdout period}
\includegraphics[width=\textwidth]{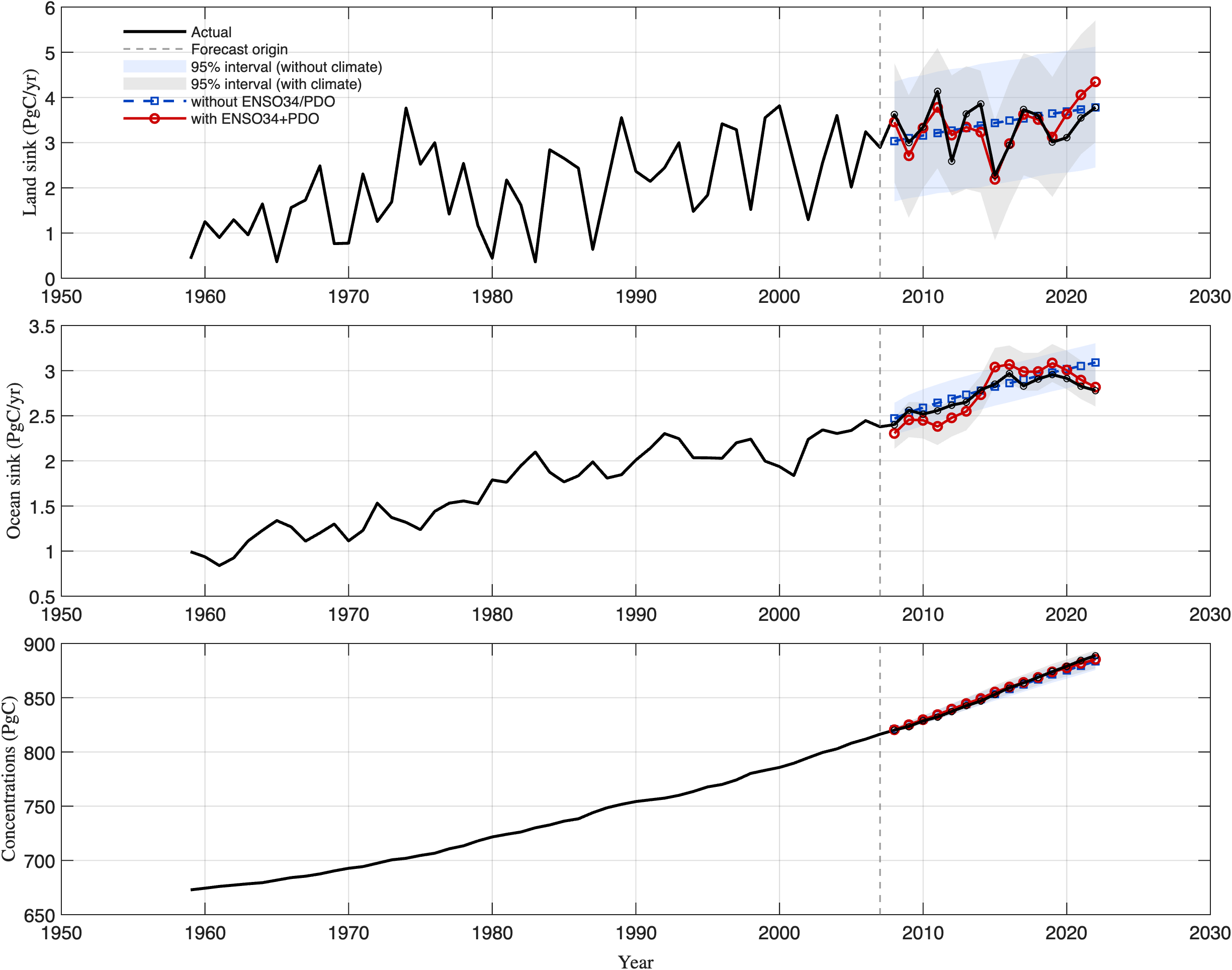}
{\footnotesize Notes: Parameters are estimated on the pre-holdout sample. The red line shows conditional projections from the restricted model with~$Z_t$, conditional on realized future emissions and realized future~$Z_t$. The blue dashed line shows the corresponding projections from the restricted model without~$Z_t$, conditional on realized future emissions only. The shaded bands are the 95\% predictive intervals, conditional on the realized exogenous paths.}
\label{fig:validation_reestimated}
\end{figure}

Figure~\ref{fig:validation_reestimated} summarizes the validation exercise.

Two features stand out. First, both versions of the restricted model reproduce the holdout paths of the land sink, ocean sink, and atmospheric concentrations quite well. This is not a mechanical implication of conditioning on emissions. Even with the realized emissions path imposed, the model must still map emissions into land uptake, ocean uptake, and atmospheric accumulation through the estimated adjustment dynamics and carbon-budget restrictions. The close fit across all three variables therefore provides supportive validation evidence for the structural mapping embodied in the model: the same restrictions used for the forward-looking projections also organize the final part of the historical record when that period is treated as out of sample.

Second, the climate-augmented specification is most informative for the sink variables. The model with $Z_t$ captures more of the year-to-year variation in the land sink than the version without climate covariates. The same pattern is present, though more modest, for the ocean sink. This is consistent with the interpretation of the climate block: conditional on emissions, stationary climate variation should primarily help explain short-run fluctuations in the sinks.

Overall, the validation evidence suggests that the restricted model can reproduce the final part of the historical record under the same conditional information set used in the scenario projections. The gains from including $Z_t$ are concentrated in the sink equations, where climate variability is expected to matter most.

\subsection{Forecasting}
\label{app:forecasting}
\label{app:forecasting_goal}

The forecasting exercise assesses whether the restricted structural model delivers competitive pseudo-out-of-sample forecasts relative to less structured alternatives under a common conditional information set. This provides a historical check on the model's use for conditional forward-looking projections. Because all models are conditioned on realized future emissions and realized future climate covariates, the exercise is best interpreted as a conditional predictive comparison rather than an unconditional forecasting contest.

We conduct a recursive pseudo-out-of-sample forecasting exercise for horizons $h=1,2,\ldots,10$. We use a balanced evaluation sample across horizons: the first forecast target is year $t=2000$ for each horizon, so that the set of target years is the same for $h=1,\ldots,10$. This choice keeps the realized evaluation years fixed across horizons while still leaving a reasonably long initial estimation sample for the longer-horizon forecasts. With the current sample ending in 2022, this gives $23$ out-of-sample forecast errors for each variable, model, and horizon. The evaluation sample is therefore informative but still modest in size, especially for formal predictive-accuracy tests. At each forecast origin, each model is estimated using only the sample available through that year. The models then produce $h$-step-ahead forecasts of $S_t^L$, $S_t^O$, and~$C_t$. Forecasts are conditional on the realized future emissions path and on the realized future path of the climate covariates~$Z_t$. Thus, the exercise evaluates each model's ability to map a given emissions and climate path into conditional forecasts of sinks and atmospheric concentrations.

For each variable and horizon, we report root mean squared forecast errors (RMSE) and mean absolute forecast errors~(MAE). We also report the results of two-sided \citet[GW]{GW2006} tests, implemented with a constant instrument, for pairwise comparisons between each benchmark model and the restricted model. For RMSE comparisons, the GW tests are based on squared-loss differentials; for MAE comparisons, they are based on absolute-loss differentials.

\subsubsection{Models}
\label{app:forecasting_models}

We compare four forecasting models.

\paragraph{Restricted structural VECM (SVECM).}
This is the restricted structural model from the main text, with the carbon-budget restrictions and the climate covariates~$Z_t$. It is the main model of interest.

\paragraph{Unrestricted VECM.}
The first benchmark is an unrestricted VECM for $Y_t = [S^L_t, S^O_t, E_t, C_t]'$ with $r=3$, $k=1$, an unrestricted constant, and a restricted trend, and includes the same stationary climate covariates. These choices mirror the unrestricted specification selected in the main-text cointegration and lag-length analysis. Unlike the restricted model, the cointegrating relations, adjustment coefficients, and short-run dynamics are unrestricted. In the conditional forecasting exercise, the realized emissions path is imposed by conditioning the joint VECM forecast on the observed future path of~$E_t$.

\paragraph{VARX in differences.}
The second benchmark is a VARX in first differences for~$Y_t$. It includes the lagged first differences, realized emissions growth, and the same climate covariates,~$Z_t$. This benchmark removes the error-correction structure and treats short-run dynamics in differences as the relevant forecasting object.

\paragraph{Univariate ARX in differences.}
The third benchmark consists of separate univariate ARX models for $\Delta S_t^L$, $\Delta S_t^O$, and~$\Delta C_t$. Each equation uses its own lagged first difference, realized emissions growth, and the same climate covariates. This benchmark removes both the error-correction structure and cross-equation interactions.

All models are evaluated under the same recursive estimation scheme, conditioning information, and forecast targets. Differences in predictive performance therefore reflect differences in model structure rather than differences in the information set.

\subsubsection{Results}
\label{app:forecasting_results}

\begin{figure}[th]
\caption{Pseudo-out-of-sample conditional forecasts, horizon $h=1$, recursive estimation}
\includegraphics[width=\textwidth]{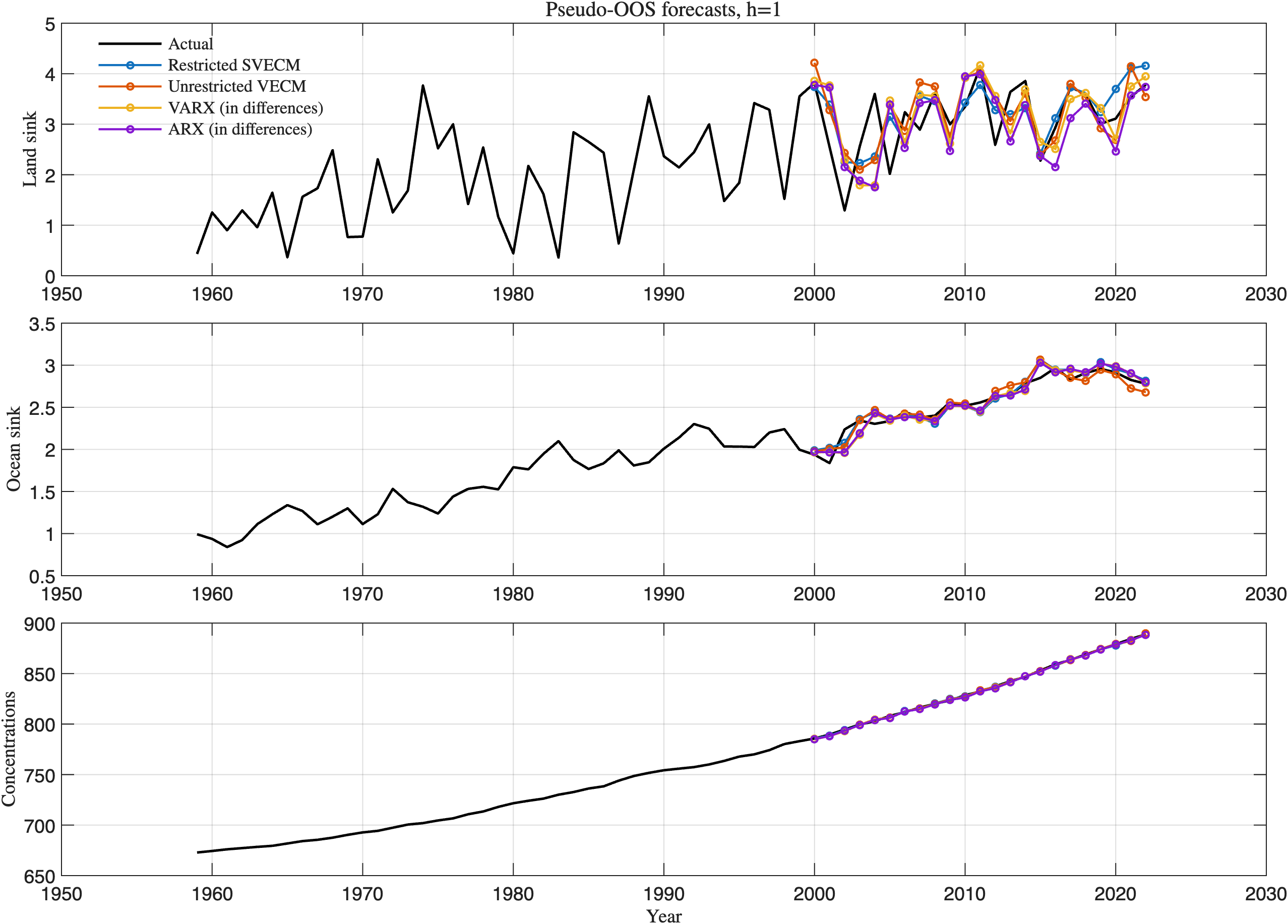}
{\footnotesize Notes: All model forecasts are conditional on realized future emissions and realized future~$Z_t$. The figure overlays the restricted structural VECM and benchmark models on the realized paths of the land sink, ocean sink, and atmospheric concentrations.}
\label{fig:forecast_h1}
\end{figure}

\begin{figure}[th]
\caption{Pseudo-out-of-sample conditional forecasts, horizon $h=10$, recursive estimation}
\includegraphics[width=\textwidth]{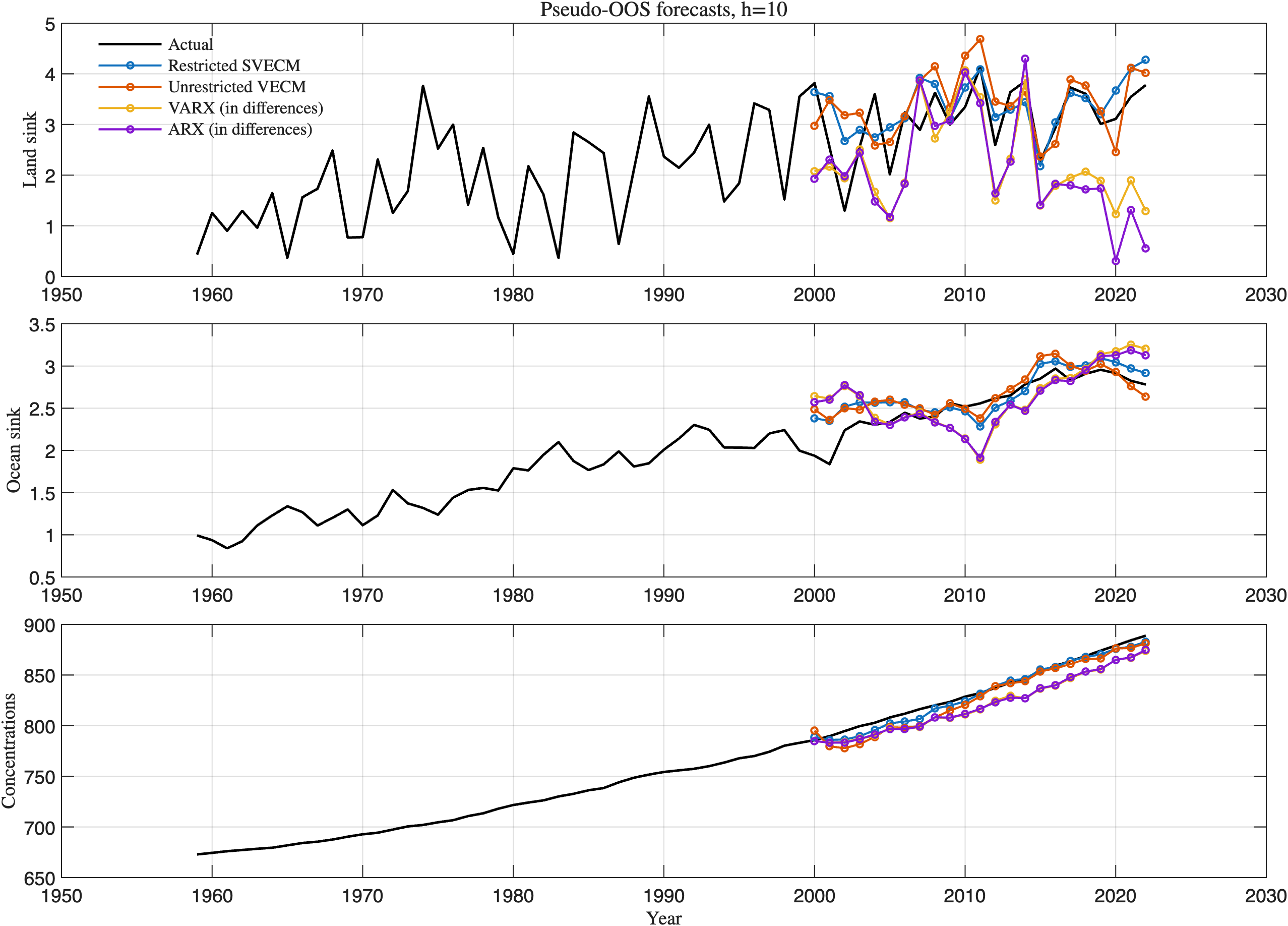}
{\footnotesize Notes: All model forecasts are conditional on realized future emissions and realized future~$Z_t$. The figure overlays the restricted structural VECM and benchmark models on the realized paths of the land sink, ocean sink, and atmospheric concentrations.}
\label{fig:forecast_h10}
\end{figure}

Figures~\ref{fig:forecast_h1} and~\ref{fig:forecast_h10} illustrate the pseudo-out-of-sample forecasts at short ($h=1$) and long ($h=10$) horizons, while Tables~\ref{tab:forecast_rmse} and~\ref{tab:forecast_mae} report the full forecast comparison.

The main conclusions are as follows.

\paragraph{Land sink.}
For the land sink, the restricted structural VECM delivers the lowest RMSE and MAE at all horizons. The gains are largest relative to the VARX and univariate ARX benchmarks, but the restricted model also improves on the unrestricted VECM. The GW tests provide evidence in favor of the restricted model against the difference-only benchmarks at essentially all horizons. Relative to the unrestricted VECM, the evidence is more moderate but generally supportive.

\paragraph{Ocean sink.}
For the ocean sink, the restricted structural VECM performs best at the shortest horizons ($h=1,2$). Under RMSE, it is also slightly more accurate at $h=10$, whereas the unrestricted VECM is more accurate at horizons $h=3,\ldots,9$. Under MAE, the unrestricted VECM is more accurate at horizons $h=3,\ldots,10$. Both error-correction models perform substantially better than the VARX and univariate ARX benchmarks. This pattern suggests that error-correction dynamics are important for the ocean sink, but that the restrictions imposed by the preferred structural specification are less uniformly beneficial for this variable than for the land sink and atmospheric concentrations.

\paragraph{Atmospheric concentrations.}
For atmospheric concentrations, the restricted structural VECM performs strongly relative to all three benchmarks. The advantage is economically large and grows with the forecast horizon, especially relative to the VARX and univariate ARX specifications. This pattern is consistent with the value of imposing the carbon-budget structure. Conditional on emissions, forecasts of $C_t$ depend on how the model allocates carbon between the atmosphere, land sink, and ocean sink. The restricted structural VECM imposes this accounting structure directly, whereas the reduced-form benchmarks approximate it statistically. The unrestricted VECM is the closest benchmark, but it generally produces larger forecast errors for~$C_t$. The GW tests support the restricted model at several horizons, although the formal evidence against the unrestricted VECM is less uniform than the magnitude of the loss differences alone might suggest.

Taken together, the pseudo-out-of-sample evidence supports the use of the restricted structural model for conditional forward-looking analysis. The strongest gains are for atmospheric concentrations and the land sink. The concentration results are particularly informative because they suggest that the carbon-budget restriction improves predictive performance, rather than merely providing an interpretable accounting framework. For the ocean sink, the restricted model remains competitive, while the unrestricted VECM provides a close and sometimes superior benchmark at medium and longer horizons.

\begin{table}[th]
\caption{Pseudo-out-of-sample forecast comparison: RMSE}
\label{tab:forecast_rmse}
\vskip -8pt
\begin{tabular*}{\linewidth}{@{\extracolsep{\fill}}lrcccc}
\hline\hline
Variable & $h$ & Restricted model & Unrestricted VECM & VARX & ARX \\
\hline
Land sink & 1 & \textbf{0.569} & 0.634$^{*}$ & 0.726$^{**}$ & 0.746$^{**}$ \\
 & 2 & \textbf{0.557} & 0.642 & 0.820$^{*}$ & 0.846$^{**}$ \\
 & 3 & \textbf{0.529} & 0.637 & 0.775$^{**}$ & 0.918$^{***}$ \\
 & 4 & \textbf{0.543} & 0.628 & 0.812$^{**}$ & 1.012$^{***}$ \\
 & 5 & \textbf{0.561} & 0.631 & 0.906$^{***}$ & 1.154$^{***}$ \\
 & 6 & \textbf{0.563} & 0.653$^{*}$ & 0.996$^{***}$ & 1.265$^{***}$ \\
 & 7 & \textbf{0.532} & 0.671$^{*}$ & 0.880$^{*}$ & 1.194 \\
 & 8 & \textbf{0.599} & 0.776$^{*}$ & 1.191$^{***}$ & 1.442$^{**}$ \\
 & 9 & \textbf{0.584} & 0.813$^{**}$ & 1.227$^{***}$ & 1.457$^{**}$ \\
 & 10 & \textbf{0.570} & 0.710$^{**}$ & 1.268$^{***}$ & 1.487$^{**}$ \\
\hline
Ocean sink & 1 & \textbf{0.089} & 0.097 & 0.101 & 0.099 \\
 & 2 & \textbf{0.110} & 0.115 & 0.132 & 0.127 \\
 & 3 & 0.122 & \textbf{0.121} & 0.165$^{**}$ & 0.158$^{*}$ \\
 & 4 & 0.131 & \textbf{0.126} & 0.189$^{***}$ & 0.181$^{**}$ \\
 & 5 & 0.143 & \textbf{0.128} & 0.202$^{**}$ & 0.189$^{**}$ \\
 & 6 & 0.153 & \textbf{0.128}$^{**}$ & 0.221$^{**}$ & 0.204$^{*}$ \\
 & 7 & 0.168 & \textbf{0.140}$^{***}$ & 0.265$^{***}$ & 0.246$^{***}$ \\
 & 8 & 0.181 & \textbf{0.158}$^{***}$ & 0.309$^{***}$ & 0.294$^{***}$ \\
 & 9 & 0.190 & \textbf{0.179}$^{**}$ & 0.328$^{***}$ & 0.317$^{***}$ \\
 & 10 & \textbf{0.208} & 0.212 & 0.352$^{***}$ & 0.335$^{***}$ \\
\hline
Conc. & 1 & \textbf{0.852} & 1.076$^{*}$ & 1.145$^{**}$ & 1.111$^{*}$ \\
 & 2 & \textbf{1.269} & 1.741 & 2.112$^{***}$ & 2.152$^{***}$ \\
 & 3 & \textbf{1.772} & 2.745 & 3.494$^{***}$ & 3.585$^{***}$ \\
 & 4 & \textbf{2.326} & 3.814 & 5.101$^{***}$ & 5.230$^{***}$ \\
 & 5 & \textbf{2.757} & 4.684 & 6.606$^{***}$ & 6.766$^{***}$ \\
 & 6 & \textbf{3.359} & 5.610 & 8.254$^{***}$ & 8.391$^{***}$ \\
 & 7 & \textbf{3.892} & 6.583 & 9.934$^{***}$ & 10.034$^{***}$ \\
 & 8 & \textbf{4.384} & 7.585$^{*}$ & 11.589$^{***}$ & 11.671$^{***}$ \\
 & 9 & \textbf{4.700} & 8.497$^{*}$ & 13.154$^{***}$ & 13.165$^{***}$ \\
 & 10 & \textbf{5.064} & 9.467$^{**}$ & 14.816$^{***}$ & 14.712$^{***}$ \\
\hline
\end{tabular*}
\vskip 4pt
{\footnotesize Notes: Entries are recursive pseudo-out-of-sample RMSE values. The evaluation sample is balanced across horizons, with first forecast target year $t=2000$ for all horizons. Bold numbers mark the lowest loss in each row. Asterisks attached to benchmark entries indicate rejection of the corresponding GW conditional predictive ability test against the restricted model; the null is equal conditional predictive ability, the GW test uses squared-loss differentials for RMSE, and $^*: p<0.10$, $^{**}: p<0.05$, $^{***}: p<0.01$.}
\end{table}

\begin{table}[th]
\caption{Pseudo-out-of-sample forecast comparison: MAE}
\label{tab:forecast_mae}
\vskip -8pt
\begin{tabular*}{\linewidth}{@{\extracolsep{\fill}}lrcccc}
\hline\hline
Variable & $h$ & Restricted model & Unrestricted VECM & VARX & ARX \\
\hline
Land sink & 1 & \textbf{0.455} & 0.496 & 0.553$^{*}$ & 0.584$^{**}$ \\
 & 2 & \textbf{0.418} & 0.492 & 0.567$^{*}$ & 0.622$^{**}$ \\
 & 3 & \textbf{0.417} & 0.496 & 0.582$^{*}$ & 0.787$^{***}$ \\
 & 4 & \textbf{0.424} & 0.505 & 0.603 & 0.843$^{***}$ \\
 & 5 & \textbf{0.408} & 0.496 & 0.701$^{***}$ & 0.953$^{***}$ \\
 & 6 & \textbf{0.442} & 0.533 & 0.836$^{***}$ & 1.101$^{***}$ \\
 & 7 & \textbf{0.414} & 0.547$^{*}$ & 0.711 & 0.928 \\
 & 8 & \textbf{0.452} & 0.590$^{**}$ & 1.108$^{***}$ & 1.286$^{***}$ \\
 & 9 & \textbf{0.433} & 0.607$^{***}$ & 0.980$^{**}$ & 1.142$^{**}$ \\
 & 10 & \textbf{0.438} & 0.576$^{***}$ & 1.103$^{***}$ & 1.240$^{**}$ \\
\hline
Ocean sink & 1 & \textbf{0.065} & 0.072 & 0.073 & 0.074 \\
 & 2 & \textbf{0.092} & 0.095 & 0.108 & 0.103 \\
 & 3 & 0.099 & \textbf{0.094} & 0.143$^{**}$ & 0.134$^{*}$ \\
 & 4 & 0.113 & \textbf{0.096} & 0.171$^{***}$ & 0.160$^{***}$ \\
 & 5 & 0.127 & \textbf{0.100}$^{**}$ & 0.176$^{**}$ & 0.164$^{*}$ \\
 & 6 & 0.134 & \textbf{0.092}$^{***}$ & 0.189$^{*}$ & 0.176 \\
 & 7 & 0.147 & \textbf{0.106}$^{***}$ & 0.217$^{**}$ & 0.200$^{**}$ \\
 & 8 & 0.158 & \textbf{0.118}$^{***}$ & 0.250$^{***}$ & 0.235$^{***}$ \\
 & 9 & 0.164 & \textbf{0.131}$^{***}$ & 0.255$^{***}$ & 0.245$^{***}$ \\
 & 10 & 0.171 & \textbf{0.153}$^{*}$ & 0.273$^{***}$ & 0.258$^{***}$ \\
\hline
Conc.& 1 & \textbf{0.746} & 0.895 & 0.989$^{**}$ & 0.938 \\
 & 2 & \textbf{1.079} & 1.390 & 1.767$^{***}$ & 1.773$^{***}$ \\
 & 3 & \textbf{1.455} & 1.969 & 3.040$^{***}$ & 3.165$^{***}$ \\
 & 4 & \textbf{2.015} & 2.841 & 4.516$^{***}$ & 4.786$^{***}$ \\
 & 5 & \textbf{2.265} & 3.334 & 6.046$^{***}$ & 6.374$^{***}$ \\
 & 6 & \textbf{2.814} & 4.241 & 7.711$^{***}$ & 7.998$^{***}$ \\
 & 7 & \textbf{3.328} & 5.172 & 9.512$^{***}$ & 9.733$^{***}$ \\
 & 8 & \textbf{3.659} & 5.927$^{*}$ & 11.195$^{***}$ & 11.392$^{***}$ \\
 & 9 & \textbf{3.908} & 6.847$^{**}$ & 12.682$^{***}$ & 12.792$^{***}$ \\
 & 10 & \textbf{4.164} & 7.829$^{***}$ & 14.205$^{***}$ & 14.141$^{***}$ \\
\hline
\end{tabular*}
\vskip 4pt
{\footnotesize Notes: Entries are recursive pseudo-out-of-sample MAE values. The evaluation sample is balanced across horizons, with first forecast target year $t=2000$ for all horizons. Bold numbers mark the lowest loss in each row. Asterisks attached to benchmark entries indicate rejection of the corresponding GW conditional predictive ability test against the restricted model; the null is equal conditional predictive ability, the GW test uses absolute-loss differentials for MAE, and $^*: p<0.10$, $^{**}: p<0.05$, $^{***}: p<0.01$.}
\end{table}

\newpage
\clearpage

\bibliographystyle{apalike}
\bibliography{mme_gcb_model}

@article{GG1995,
    author = {Gonzalo, J. and Granger, C.},
    title = {Estimation of common long-memory components in cointegrated systems},
    journal = {Journal of Business \& Economic Statistics},
    volume = {13},
    pages = {27--35},
    year = {1995}
}

@article{GW2006,
    author = {Giacomini, R. and White, H.},
    title = {Tests of conditional predictive ability},
    journal = {Econometrica},
    volume = {74},
    pages = {1545--1578},
    year = {2006}
}

@article{CRT2012,
    author = {Cavaliere, G. and Rahbek, A. and Taylor, A. M. R.},
    title = {Bootstrap determination of the co-integration rank in vector autoregressive models},
    journal = {Econometrica},
    volume = {80},
    pages = {1721--1740},
    year = {2012}
}

@article{Schmith2012,
    author = {Schmith, T. and Johansen, S. and Thejll, P.},
    title = {Statistical analysis of global surface temperature and sea level using cointegration methods},
    journal = {Journal of Climate},
    volume = {25},
    pages = {7822--7833},
    year = {2012}
}

@article{Pretis2020,
    author = {Pretis, Felix},
    title = {Econometric modelling of climate systems: The equivalence of energy balance models and cointegrated vector autoregressions},
    journal = {Journal of Econometrics},
    volume = {214},
    pages = {256--273},
    year = {2020}
}

@techreport{Castle2024,
    author = {Castle, J. L. and Doornik, J. A. and Hendry, D. F.},
    title = {Forecasting climate change using a multivariate cointegrated system},
    year = {2024},
	type = {Working paper},
	institution = {Oxford University}
}

@article{boswijk2004,
author = {Boswijk, H. Peter and Doornik, Jurgen A.},
title = {Identifying, estimating and testing restricted cointegrated systems: An overview},
journal = {Statistica Neerlandica},
volume = {58},
pages = {440--465},
year = {2004}
}

@book{Johansen1995,
    author = {Johansen, Søren},
    title = "{Likelihood-Based Inference in Cointegrated Vector Autoregressive Models}",
    publisher = {Oxford University Press},
    year = {1995},
    isbn = {9780198774501},
    doi = {10.1093/0198774508.001.0001},
    url = {https://doi.org/10.1093/0198774508.001.0001}
}

@book{Juselius2006,
    author={Juselius, Katarina},
    title={The Cointegrated VAR Model: Methodology and Applications},
    publisher={Oxford University Press},
    year={2006},
    isbn={9780199285662}
}

@article{BHK2023,
    author = {Bennedsen, Mikkel and Hillebrand, Eric and Koopman, Siem Jan},
    title = {A multivariate dynamic statistical model of the global carbon budget 1959--2020},
    journal = {Journal of the Royal Statistical Society Series A: Statistics in Society},
    volume = {186},
    number = {1},
    pages = {20--42},
    year = {2023},
    month = {01},
    issn = {0964-1998},
    doi = {10.1093/jrsssa/qnac014},
    url = {https://doi.org/10.1093/jrsssa/qnac014},
    eprint = {https://academic.oup.com/jrsssa/article-pdf/186/1/20/49348925/qnac014.pdf}
}

@article{BHK2024,
    author = {Bennedsen, Mikkel and Hillebrand, Eric and Koopman, Siem Jan},
    title = {A regression-based approach to the {CO$_2$} airborne fraction},
    journal = {Nature Communications},
    volume = {15},
    number = {8507},
    pages = {1--9},
    year = {2024},
    doi = {10.1038/s41467-024-52728-1},
    url = {https://doi.org/10.1038/s41467-024-52728-1}
}

@misc{Enting1993,
	author = {Enting, I.G. and Lassey, K.R.},
	howpublished = {{CSIRO Division of Atmospheric Research Technical Paper no. 27}, http://www.cmar.csiro.au/e-print/open/enting\_2000e.pdf},
	title = {{Projections of Future CO$_2$}},
	year = {1993}}

@article{Parkinson1998,
	author = {Parkinson, Stuart and Young, Peter},
	journal = {Climate Research},
	volume = {9},
	pages = {157--174},
	title = {Uncertainty and sensitivity in global carbon cycle modeling},
	year = {1998}}

@article{LeQuere2007,
	author = {Le Qu{\'e}r{\'e}, Corinne and R{\"o}denbeck, Christian and Buitenhuis, Erik T. and Conway, Thomas J. and Langenfelds, Ray and Gomez, Antony and Labuschagne, Casper and Ramonet, Michel and Nakazawa, Takakiyo and Metzl, Nicolas and Gillett, Nathan and Heimann, Martin},
	journal = {Science},
	number = {5832},
	pages = {1735--1738},
	title = {Saturation of the {Southern} {Ocean} {CO$_2$} Sink Due to Recent Climate Change},
	volume = {316},
	year = {2007}}

@incollection{Canadell2007Saturation,
	address = {Berlin, Heidelberg},
	author = {Canadell, Josep G. and Pataki, Diane E. and Gifford, Roger and Houghton, Richard A. and Luo, Yiqi and Raupach, Michael R. and Smith, Pete and Steffen, Will},
	booktitle = {Terrestrial Ecosystems in a Changing World},
	date = {2007//},
	doi = {10.1007/978-3-540-32730-1{\_}6},
	editor = {Canadell, Josep G. and Pataki, Diane E. and Pitelka, Louis F.},
	id = {Canadell2007},
	isbn = {978-3-540-32730-1},
	pages = {59--78},
	publisher = {Springer Berlin Heidelberg},
	title = {Saturation of the Terrestrial Carbon Sink},
	url = {https://doi.org/10.1007/978-3-540-32730-1_6},
	year = {2007}}

@incollection{AR6chapter5,
	author = {Canadell and J.G. and P.M.S. Monteiro and M.H. Costa and L. Cotrim da Cunha and P.M. Cox and A.V. Eliseev and S. Henson and M. Ishii and S. Jaccard and C. Koven and A. Lohila and P.K. Patra and S. Piao and J. Rogelj and S. Syampungani and S. Zaehle and K. Zickfeld},
	booktitle = {Climate Change 2021: The Physical Science Basis. Contribution of Working Group I to the Sixth Assessment Report of the Intergovernmental Panel on Climate Change},
	editor = {Masson-Delmotte, V. and others},
	publisher = {Cambridge University Press},
	title = {Global Carbon and other Biogeochemical Cycles and Feedbacks},
	year = {2021}}

@Article{GCB2023,
AUTHOR = {Friedlingstein, P. and O'Sullivan, M. and Jones, M. W. and Andrew, R. M. and Bakker, D. C. E. and Hauck, J. and Landsch\"utzer, P. and Le Qu\'er\'e, C. and Luijkx, I. T. and Peters, G. P. and Peters, W. and Pongratz, J. and Schwingshackl, C. and Sitch, S. and Canadell, J. G. and Ciais, P. and Jackson, R. B. and Alin, S. R. and Anthoni, P. and Barbero, L. and Bates, N. R. and Becker, M. and Bellouin, N. and Decharme, B. and Bopp, L. and Brasika, I. B. M. and Cadule, P. and Chamberlain, M. A. and Chandra, N. and Chau, T.-T.-T. and Chevallier, F. and Chini, L. P. and Cronin, M. and Dou, X. and Enyo, K. and Evans, W. and Falk, S. and Feely, R. A. and Feng, L. and Ford, D. J. and Gasser, T. and Ghattas, J. and Gkritzalis, T. and Grassi, G. and Gregor, L. and Gruber, N. and G\"urses, \"O. and Harris, I. and Hefner, M. and Heinke, J. and Houghton, R. A. and Hurtt, G. C. and Iida, Y. and Ilyina, T. and Jacobson, A. R. and Jain, A. and Jarn\'{\i}kov\'a, T. and Jersild, A. and Jiang, F. and Jin, Z. and Joos, F. and Kato, E. and Keeling, R. F. and Kennedy, D. and Klein Goldewijk, K. and Knauer, J. and Korsbakken, J. I. and K\"ortzinger, A. and Lan, X. and Lef\`evre, N. and Li, H. and Liu, J. and Liu, Z. and Ma, L. and Marland, G. and Mayot, N. and McGuire, P. C. and McKinley, G. A. and Meyer, G. and Morgan, E. J. and Munro, D. R. and Nakaoka, S.-I. and Niwa, Y. and O'Brien, K. M. and Olsen, A. and Omar, A. M. and Ono, T. and Paulsen, M. and Pierrot, D. and Pocock, K. and Poulter, B. and Powis, C. M. and Rehder, G. and Resplandy, L. and Robertson, E. and R\"odenbeck, C. and Rosan, T. M. and Schwinger, J. and S\'ef\'erian, R. and Smallman, T. L. and Smith, S. M. and Sospedra-Alfonso, R. and Sun, Q. and Sutton, A. J. and Sweeney, C. and Takao, S. and Tans, P. P. and Tian, H. and Tilbrook, B. and Tsujino, H. and Tubiello, F. and van der Werf, G. R. and van Ooijen, E. and Wanninkhof, R. and Watanabe, M. and Wimart-Rousseau, C. and Yang, D. and Yang, X. and Yuan, W. and Yue, X. and Zaehle, S. and Zeng, J. and Zheng, B.},
TITLE = {Global Carbon Budget 2023},
JOURNAL = {Earth System Science Data},
VOLUME = {15},
YEAR = {2023},
NUMBER = {12},
PAGES = {5301--5369},
URL = {https://essd.copernicus.org/articles/15/5301/2023/},
DOI = {10.5194/essd-15-5301-2023}
}

@article{Bennedsen2021,
	author = {M. Bennedsen},
	doi = {10.1007/s10584-021-03123-y},
	journal = {Climatic Change},
	number = {21},
	pages = {1--19},
	title = {Designing a statistical procedure for monitoring global carbon dioxide emissions},
	volume = {166},
	year = {2021}}

@article{Peters2017,
	author = {Peters, Glen P. and Le Qu{\'e}r{\'e}, Corinne and Andrew, Robbie M. and Canadell, Josep G. and Friedlingstein, Pierre and Ilyina, Tatiana and Jackson, Robert B. and Joos, Fortunat and Korsbakken, Jan Ivar and McKinley, Galen A. and Sitch, Stephen and Tans, Pieter},
	journal = {Nature Climate Change},
	number = {12},
	pages = {848--850},
	title = {Towards real-time verification of {CO$_2$} emissions},
	volume = {7},
	year = {2017}}

@article{feely1999influence,
	author = {Feely, R. A. and Wanninkhof, R. and Takahashi, T. and Tans, P.},
	journal = {Nature},
	number = {6728},
	pages = {597--601},
	title = {Influence of {E}l {N}i\~no on the equatorial {Pacific} contribution to atmospheric {CO$_2$} accumulation},
	volume = {398},
	year = {1999}}

@article{haverd2018new,
	author = {Haverd, V. and Smith, B. and Nieradzik, L. and Briggs, P.R. and Woodgate, W. and Trudinger, C.M. and Canadell, J.G. and Cuntz, M.},
	journal = {Geoscientific Model Development},
	pages = {2995--3026},
	title = {A new version of the {CABLE} land surface model (subversion revision r4601) incorporating land use and land cover change, woody vegetation demography, and a novel optimisation-based approach to plant coordination of photosynthesis},
	volume = {11},
	year = {2018}}

@article{Knorr2009,
	author = {W. Knorr},
	journal = {Geophysical Research Letters},
	title = {Is the airborne fraction of anthropogenic {CO$_2$} emissions increasing?},
	volume = {36},
	year = {2009}}

@article{Raupach2008,
	author = {M. R. Raupach and J. G. Canadell and C. Le Qu\'er\'e},
	journal = {Biogeosciences},
	pages = {1601--1613},
	title = {Anthropogenic and biophysical contributions to increasing atmospheric {CO$_2$} growth rate and airborne fraction},
	volume = {5},
	year = {2008}}

@article{LeQuere2009,
	author = {Le Qu{\'e}r{\'e}, C. and Raupach, M.R. and Canadell, J.G. and Marland, G. and Bopp, L. and Ciais, P. and Conway, T.J. and Doney, S.C. and Feely, R.A. and Foster, P. and Friedlingstein, P. and Gurney, K. and Houghton, R.A. and House, J.I. and Huntingford, C. and Levy, P.E. and Lomas, M.R. and Majkut, J. and Metzl, N. and Ometto, J.P. and Peters, G.P. and Prentice, I.C. and Randerson, J.T. and Running, S.W. and Sarmiento, J.L. and Schuster, U. and Sitch, S. and Takahashi, T. and Viovy, N. and van der Werf, G.R. and Woodward, F.I.},
	journal = {Nature Geoscience},
	pages = {831--836},
	title = {Trends in the sources and sinks of carbon dioxide},
	volume = {2},
	year = {2009}}

@article{BHK2020,
        author = {Bennedsen, Mikkel and Hillebrand, Eric and Koopman, Siem Jan},
	doi = {https://doi.org/10.1016/j.eneco.2021.105118},
	journal = {Energy Economics},
	pages = {105118},
	title = {Modeling, forecasting, and nowcasting {U.S.} {CO$_2$} emissions using many macroeconomic predictors},
	volume = {96},
	year = {2021}}

@article{BHK2019,
        author = {Bennedsen, Mikkel and Hillebrand, Eric and Koopman, Siem Jan},
	journal = {Biogeosciences},
	number = {18},
	pages = {3651--3663},
	title = {Trend analysis of the airborne fraction and sink rate of anthropogenically released {CO$_2$}},
	volume = {16},
	year = {2019}}

@article{Raupach2014,
	author = {Raupach, M. R. and Gloor, M. and Sarmiento, J. L. and Canadell, J. G. and Fr\"olicher, T. L. and Gasser, T. and Houghton, R. A. and Le Qu\'er\'e, C. and Trudinger, C. M.},
	doi = {10.5194/bg-11-3453-2014},
	journal = {Biogeosciences},
	number = {13},
	pages = {3453--3475},
	title = {The declining uptake rate of atmospheric {CO}$_{2}$ by land and ocean sinks},
	url = {https://www.biogeosciences.net/11/3453/2014/},
	volume = {11},
	year = {2014}}

@article{Gloor2010,
	author = {M. Gloor and J. L. Sarmiento and N. Gruber},
	journal = {Atmospheric Chemistry and Physics},
	pages = {7739--7751},
	title = {What can be learned about carbon cycle climate feedbacks from the {CO$_2$} airborne fraction?},
	volume = {10},
	year = {2010}}

@incollection{BK1973,
	author = {R. Bacastow and C. D. Keeling},
	booktitle = {Carbon and the Biosphere Conference Proceedings},
    address = {Upton, New York, USA},
	pages = {86--135},
	publisher = {Brookhaven Symposia in Biology},
	title = {Atmospheric carbon dioxide and radiocarbon in the natural cycle: {II}. Changes from {A. D.} 1700 to 2070 as deduced from a geochemical model},
	year = {1973}}

@incollection{gifford1993implications,
	author = {Gifford, R.M.},
	booktitle = {The Global Carbon Cycle},
	editor = {Heimann, M.},
	pages = {159--199},
	publisher = {Springer},
	title = {Implications of {CO$_2$} effects on vegetation for the global carbon budget},
	year = {1993}}

@incollection{IPCC2001_3rd,
	author = {Prentice, I.C. and Farquhar, G.D. and Fasham,M.J.R. and Goulden, M.L. and Heimann, M. and Jaramillo, V.J. and Kheshgi, H.S. and Le Quéré, C. and Scholes, R.J. and Wallace, D.W.R.},
	publisher = {Cambridge University Press},
	title = {{The carbon cycle and atmospheric carbon dioxide}},
    booktitle = {{Climate Change
2001: The Scientific Basis Contribution of Working Group I to the Third Assessment Report of the Intergovernmental Panel on Climate Change}},
	year = {2001}}

@article{MAGICC,
	author = {Meinshausen, M. and Raper, S. C. B. and Wigley, T. M. L.},
	journal = {Atmospheric Chemistry and Physics},
	number = {4},
	pages = {1417--1456},
	title = {Emulating coupled atmosphere-ocean and carbon cycle models with a simpler model, {MAGICC}6 -- {Part} 1: {Model} description and calibration},
	volume = {11},
	year = {2011}}

@article{Friedlingstein2015,
	author = {P. Friedlingstein},
	journal = {Philosophical Transactions of the Royal Society A},
	pages = {20140421},
	title = {Carbon cycle feedbacks and future climate change},
	volume = {373},
	year = {2015}}

@Article{oneill2016,
AUTHOR = {O'Neill, B. C. and Tebaldi, C. and van Vuuren, D. P. and Eyring, V. and Friedlingstein, P. and Hurtt, G. and Knutti, R. and Kriegler, E. and Lamarque, J.-F. and Lowe, J. and Meehl, G. A. and Moss, R. and Riahi, K. and Sanderson, B. M.},
TITLE = {The scenario model intercomparison project {(ScenarioMIP)} for {CMIP6}},
JOURNAL = {Geoscientific Model Development},
VOLUME = {9},
YEAR = {2016},
NUMBER = {9},
PAGES = {3461--3482},
URL = {https://gmd.copernicus.org/articles/9/3461/2016/},
DOI = {10.5194/gmd-9-3461-2016}
}

@article{riahi2017,
title = {{The Shared Socioeconomic Pathways and their energy, land use, and greenhouse gas emissions implications: An overview}},
journal = {Global Environmental Change},
volume = {42},
pages = {153--168},
year = {2017},
issn = {0959-3780},
doi = {https://doi.org/10.1016/j.gloenvcha.2016.05.009},
url = {https://www.sciencedirect.com/science/article/pii/S0959378016300681},
author = {Keywan Riahi and Detlef P. {van Vuuren} and Elmar Kriegler and Jae Edmonds and Brian C. O’Neill and Shinichiro Fujimori and Nico Bauer and Katherine Calvin and Rob Dellink and Oliver Fricko and Wolfgang Lutz and Alexander Popp and Jesus Crespo Cuaresma and Samir KC and Marian Leimbach and Leiwen Jiang and Tom Kram and Shilpa Rao and Johannes Emmerling and Kristie Ebi and Tomoko Hasegawa and Petr Havlik and Florian Humpenöder and Lara Aleluia {Da Silva} and Steve Smith and Elke Stehfest and Valentina Bosetti and Jiyong Eom and David Gernaat and Toshihiko Masui and Joeri Rogelj and Jessica Strefler and Laurent Drouet and Volker Krey and Gunnar Luderer and Mathijs Harmsen and Kiyoshi Takahashi and Lavinia Baumstark and Jonathan C. Doelman and Mikiko Kainuma and Zbigniew Klimont and Giacomo Marangoni and Hermann Lotze-Campen and Michael Obersteiner and Andrzej Tabeau and Massimo Tavoni}
}

@article{Trenberth1997,
    author = {Trenberth, Kevin E.},
    title = {The definition of {El Ni\~no}},
    journal = {Bulletin of the American Meteorological Society},
    volume = {78},
    number = {12},
    pages = {2771--2778},
    year = {1997},
    doi = {10.1175/1520-0477(1997)078<2771:TDOENO>2.0.CO;2}
}

@article{Rayner2003,
    author = {Rayner, N. A. and Parker, D. E. and Horton, E. B. and Folland, C. K. and Alexander, L. V. and Rowell, D. P. and Kent, E. C. and Kaplan, A.},
    title = {Global analyses of sea surface temperature, sea ice, and night marine air temperature since the late nineteenth century},
    journal = {Journal of Geophysical Research: Atmospheres},
    volume = {108},
    number = {D14},
    pages = {4407},
    year = {2003},
    doi = {10.1029/2002JD002670}
}

@misc{Nino34_PSL,
    author = {{NOAA Physical Sciences Laboratory}},
    title = {{Ni\~no~3.4 SST Anomaly Index from HadISST, 1981--2010 base period}},
    howpublished = {\url{https://psl.noaa.gov/data/timeseries/month/data/nino34.long.anom.data}},
    year = {2026},
    note = {Accessed on 2026-04-27}
}

@article{Mantua1997,
    author = {Mantua, Nathan J. and Hare, Steven R. and Zhang, Yuan and Wallace, John M. and Francis, Robert C.},
    title = {A {Pacific} interdecadal climate oscillation with impacts on salmon production},
    journal = {Bulletin of the American Meteorological Society},
    volume = {78},
    number = {6},
    pages = {1069--1079},
    year = {1997},
    doi = {10.1175/1520-0477(1997)078<1069:APICOW>2.0.CO;2}
}

@article{Huang2017,
    author = {Huang, Boyin and Thorne, Peter W. and Banzon, Viva F. and Boyer, Tim and Chepurin, Gennady and Lawrimore, Jay H. and Menne, Matthew J. and Smith, Thomas M. and Vose, Russell S. and Zhang, Huai-Min},
    title = {Extended reconstructed sea surface temperature, version 5 ({ERSSTv5}): upgrades, validations, and intercomparisons},
    journal = {Journal of Climate},
    volume = {30},
    number = {20},
    pages = {8179--8205},
    year = {2017},
    doi = {10.1175/JCLI-D-16-0836.1}
}

@misc{PDO_NCEI,
    author = {{NOAA National Centers for Environmental Information}},
    title = {{Pacific Decadal Oscillation (PDO) Index, derived from ERSSTv5}},
    howpublished = {\url{https://www.ncei.noaa.gov/pub/data/cmb/ersst/v5/index/ersst.v5.pdo.dat}},
    year = {2026},
    note = {Accessed on 2026-04-27}
}

\end{document}